\documentclass[%
 aip,
 amsmath,amssymb
 reprint,%
]{revtex4-1}
\usepackage{graphicx}
\usepackage{float} 
\usepackage{dcolumn}
\usepackage{bm}
\usepackage{amsmath} \usepackage{amssymb}
\usepackage[dvipsnames]{xcolor}
\usepackage{soul}
\usepackage[utf8]{inputenc}
\usepackage[T1]{fontenc}
\usepackage{mathptmx}
\usepackage{etoolbox}


\begin{document}


\title{Phototactic Bioconvection in a Rotating Isotropic Porous Medium: Linear Stability Analysis}
\author{Sandeep Kumar}
\email{sandeepkumar1.00123@gmail.com}
\affiliation{
Department of Energy Science and Engineering, Indian Institute of Technology Bombay, Mumbai 400076, India
}%
\author{Suneet Singh}
\affiliation{
Department of Energy Science and Engineering, Indian Institute of Technology Bombay, Mumbai 400076, India
}%


\date{\today}

\begin{abstract}
This study investigates the linear stability of phototactic bioconvection in a rotating porous medium under collimated light, incorporating the effects of critical intensity, Darcy number, and Taylor number. Using a mathematical model and the MATLAB Bvp4c solver, the critical Rayleigh number and wavenumber for instability onset are identified. The results reveal that higher Darcy numbers enhance instability, increasing the wavelength of bioconvection patterns, while rotation exerts a stabilizing effect by limiting vertical motion and confining fluid dynamics to the horizontal plane. Additionally, an increase in critical intensity amplifies instability. Furthermore, the study explores the transition between oscillatory and stationary solutions, highlighting the role of rotational dynamics in altering instability modes. These findings provide novel insights into the interplay of phototaxis, rotation, and porous media, advancing the understanding of bioconvective systems with potential applications in environmental engineering, biophysics, and geophysical fluid dynamics. 

\end{abstract}

\maketitle

\begin{quotation}

\end{quotation}

\section{Introduction}

Bioconvection refers to the spontaneous formation of large-scale flow patterns in suspensions of self-propelled microorganisms, such as algae and bacteria, due to their collective motion \citep{ref1,ref3,ref7a}. These microorganisms, which are slightly denser than water, tend to swim upward, forming denser horizontal layers known as sublayers. When the density gradient across these sublayers becomes sufficiently large, the system becomes unstable, and bioconvection patterns emerge \citep{ref3, ref5}. The name "bioconvection" was first introduced by \citet{ref3}, while earlier experimental studies by \citet{ref2} laid the foundation for this phenomenon. The movement of microorganisms is driven by their behavioral response to external stimuli, collectively known as \textit{taxis}. Among these, phototaxis the directed movement of microorganisms in response to light is particularly significant. Phototactic microorganisms, such as \textit{Chlamydomonas}, \textit{Euglena}, and \textit{Volvox}, exhibit positive phototaxis (movement toward light) or negative phototaxis (movement away from light) depending on light intensity \citep{ref2,ref4,ref5, ref6}. The patterns generated by bioconvection are influenced by factors such as suspension depth, microorganism concentration, and light absorption \citep{ref7}. Specifically, variations in total light intensity $\mathbb{J}$ relative to the critical intensity $\mathbb{J}_c$ can determine whether microorganisms accumulate at the top or bottom of the suspension, altering both the shape and size of bioconvection patterns \citep{ref5,ref8}.

Bioconvection in a porous medium introduces additional complexity and is an emerging area of research with significant scientific and engineering implications \cite{kopp2023effect}. Porous media are ubiquitous in natural and engineered systems, such as soil, filtration units, and biological tissues. Understanding bioconvection in porous environments can lead to advancements in fields such as groundwater bioremediation, nutrient transport in aquifers, enhanced oil recovery, and tissue engineering \citep{rajput2024mathematical}. In such systems, microorganisms interact with fluid flow and the porous matrix, leading to novel instability mechanisms. Notably, the shading effect where algae cast shadows on cells beneath them can further modify the observed bioconvection patterns under light irradiation. Numerous studies have investigated the role of gravitactic, gyrotactic, and oxytactic microorganisms in bioconvection within porous media. Kuznetsov and Avramenko \cite{kuznetsov2001numerical} examined gravitactic bioconvection, identifying the critical permeability needed for its onset, while they also analyzed gyrotactic effects in porous layers \cite{kuznetsov2003stability}. Nield et al. \cite{nield2004onset} studied the influence of gyrotaxis at the onset of bioconvection, and Avramenko and Kuznetsov \cite{avramenko2006onset} explored vertical throughflow impacts on gyrotactic suspensions in porous fluid layers. Biswas et al. \cite{biswas2020thermo} studied the thermo-bioconvection of oxytactic microorganisms under a magnetic field in a porous medium. Kopp and Yanovsky \cite{kopp2023effect, kopp2023darcy} explored bio-thermal convection in porous layers, analyzing the effects of rotation, gravity modulation, and gyrotactic microorganisms. They further investigated thermal bioconvection under combined influences of rotation, gravity modulation, and heat sources \cite{kopp2024weakly}.

In many natural and industrial processes, rotation also plays a crucial role in fluid dynamics. Rotational effects arise in geophysical contexts, such as atmospheric circulation, ocean currents, and convection in Earth's core, as well as in astrophysical systems \citep{ref11}. Rotation fundamentally alters the stability of fluid layers by introducing Coriolis forces, which interact with the convective flows. Combining phototaxis with rotation in a porous medium creates a rich and complex system where light-induced accumulation, shading effects, and rotational dynamics interact to drive bioconvective instability. Fluid dynamics under rotation has garnered significant interest for its complexity and applications \citep{ref11}. In viscous fluids, the Bénard convection under rotation is well-studied \citep{ref12}. \citet{ref12-1} explored thermoconvective instability in ferromagnetic fluids within rotating porous media under magnetic fields, while \citet{ref12-2} investigated ferrofluid layers heated from below, showing slower stabilization than viscous fluids. \citet{ref12-3} analyzed rotational effects using Boussinesq equations, and \citet{ref12-4} highlighted boundary-dependent instabilities in fluids with conductivity gradients. Studies on rotating nanofluids, such as by \citet{ref12-5} and \citet{ref12-6}, demonstrated enhanced stability with increased rotation and magnetic strength, while \citet{waqas} analyzed Casson nanofluid bioconvection on revolving disks.

The concept of phototactic bioconvection was first introduced by Vincent and Hill \cite{ref9}, who developed a model focusing on non-scattering, absorbing suspensions. Their work investigated the stability of a suspension of phototactic microorganisms under uniform illumination from above, with the microorganisms swimming in a slightly less dense fluid. This foundational model included the effects of phototaxis and shading. Later, Ghorai and Hill \cite{ref13} extended this research by proposing a two-dimensional phototactic bioconvection model, where the stability was analyzed using a conservative finite-difference numerical approach. However, these models did not account for scattering effects. To address this limitation, Ghorai et al. \cite{ref14} advanced the phototactic bioconvection model by incorporating isotropic scattering. Their findings revealed that scattering could lead to the aggregation of microorganisms in two distinct horizontal layers at different depths, depending on specific parameter values. Building on this, Kumar \cite{kumar2023} examined the stability of suspensions and demonstrated that a rigid top surface resulted in greater system stability compared to a stress-free surface. Further research by Ghorai and Panda \cite{ref15} explored the effects of forward anisotropic scattering on phototactic bioconvection suspensions. Their computational analysis highlighted the critical role of forward scattering in influencing suspension stability. Subsequently, Panda et al. \cite{ref16} investigated how suspensions responded to both diffuse and collimated light, uncovering diverse behavioral patterns under varying conditions. Moreover, Panda et al. \cite{ref17} examined phototactic bioconvection under oblique illumination in non-scattering suspensions, identifying transitions between stable and overstable modes for different angles of incidence and parameter combinations. Kumar \cite{ref18}, extended this research by investigating the effects of oblique irradiation on isotropic scattering suspensions. Their findings indicated that variations in the angle of incidence and scattering albedo significantly influenced concentration profiles and stability. Despite these advancements, the role of rotation in phototactic bioconvection had remained unexplored until Kumar \cite{kumar2023effect} analyzed its impact on a non-scattering medium. His study illustrated that rotation exerts a stabilizing influence on the suspension. Rajput and Panda \cite{rajput2024effect} further examined the influence of diffuse flux on isotropic scattering suspensions, revealing that increased levels of diffuse flux enhance suspension stability by raising the critical Rayleigh number. However, there remains a significant gap in the literature concerning the combined effects of phototaxis and temperature gradients. Kumar and Wang \cite{kumar2024thermal,sandeep2024thermal} recently addressed this gap by studying phototactic bioconvection in non-scattering suspensions under heating from below/above and collimated irradiation. Their analysis utilized the MATLAB bvp$4$c solver to examine the system's dynamics and stability, marking a step forward in understanding these complex interactions. Recent research by \citet{rajput2024mathematical} highlights the significance of studying bioconvection in isotropic porous media illuminated by collimated light, emphasizing the role of parameters like Darcy number, critical light intensity, and swimming speed in stimulating bioconvection. 

Despite extensive studies on bioconvection, a significant research gap exists regarding light-induced bioconvection in rotating porous media. Addressing this gap is critical for enhancing our understanding of how phototaxis, rotation, and porous structures collectively influence fluid behavior. The current study aims to investigate the linear stability of a suspension of phototactic microorganisms in a rotating porous medium illuminated by vertically collimated light. By incorporating shading effects and rotational dynamics, we provide novel insights into the mechanisms driving bioconvection under realistic conditions. 

This article is structured as follows: Section \ref{mathformulation} outlines the mathematical formulation of the problem. Section \ref{Basic State} focuses on deriving the basic state solution. In Section \ref{Linear Analysis}, the linear stability analysis of the suspension is conducted, followed by the normal mode analysis presented in Section \ref{Normal Mode Analysis}. Section \ref{Numerical solution} discusses the numerical results obtained, and finally, Section \ref{Conclusion} provides a comprehensive conclusion of the study.



\begin{figure}
    \centering
    \includegraphics[width=16cm, height=11cm ]{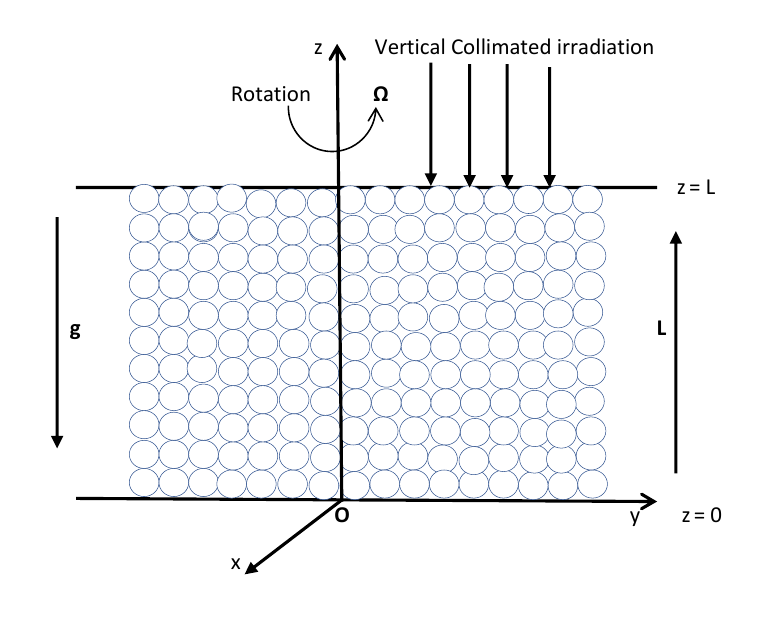}
    \caption{Geometrical view of the porous rotating medium.}
   \label{fig:recent.pdf}
 \end{figure}

\section{MATHEMATICAL FORMULATION}
\label{mathformulation}
\subsection{System Description} 

We consider a three-dimensional Cartesian coordinate system \( Oxyz \) with a finite vertical depth \( L \) and infinite horizontal width. An isotropic porous medium is confined between two boundaries: the lower boundary at \( z=0 \) (rigid) and the upper boundary at \( z=L \) (stress-free). The entire fluid system and the boundaries rotate uniformly about the \( z \)-axis with a constant angular velocity \( \boldsymbol{\Omega} = \Omega \hat{\boldsymbol{z}} \) (see Fig. \ref{fig:recent.pdf}). The system is illuminated from above by a vertically collimated light source, and the effects of oblique or diffused irradiation are ignored.

\subsection{Intensity and Radiative Heat Flux}
The intensity \( G(\boldsymbol{X}, \hat{\boldsymbol{u}}) \) at a position \( \boldsymbol{X} = (x, y, z) \) in the direction \( \hat{\boldsymbol{u}} \) is given as:
\begin{equation}
\hat{\boldsymbol{u}} = \sin{\beta_1}\cos{\beta_2}\hat{\boldsymbol{x}} + \sin{\beta_1}\sin{\beta_2}\hat{\boldsymbol{y}} + \cos{\beta_1}\hat{\boldsymbol{z}},
\end{equation}
where \( \beta_1 \) and \( \beta_2 \) are the polar and azimuthal angles. \( \hat{\boldsymbol{x}}, \hat{\boldsymbol{y}}, \) and \( \hat{\boldsymbol{z}} \) are unit vectors in the \( x \), \( y \), and \( z \) directions, respectively. The total light intensity \( \mathbb{J}(\boldsymbol{X}) \) and radiative heat flux \( \boldsymbol{f}(\boldsymbol{X}) \) at a point \( \boldsymbol{X} \) are defined as \cite{ref-modest}:
\begin{equation}
\mathbb{J}(\boldsymbol{X}) = \int_0^{4\pi} G(\boldsymbol{X}, \hat{\boldsymbol{u}}) d\omega, \quad
\boldsymbol{f}(\boldsymbol{X}) = \int_0^{4\pi} G(\boldsymbol{X}, \hat{\boldsymbol{u}}) \hat{\boldsymbol{u}} d\omega,
\end{equation}
where \( \omega \) is the solid angle.

The light intensity \( G(\boldsymbol{X}, \hat{\boldsymbol{u}}) \) in a non-scattering, absorbing medium satisfies the Radiative Transfer Equation (RTE) \cite{ref-modest,ref-chand}:
\begin{equation}
\hat{\boldsymbol{u}} \cdot \nabla G(\boldsymbol{X}, \hat{\boldsymbol{u}}) + \psi G(\boldsymbol{X}, \hat{\boldsymbol{u}}) = 0,
\end{equation}
where \( \psi \) is the absorption coefficient, proportional to the cell concentration \( n \), such that \( \psi = \iota n \). 

Boundary conditions are applied as follows:  
At the top surface \( z=L \), the incoming collimated light satisfies:
\begin{equation}
G(x, y, L, \beta_1, \beta_2) = G_0 \delta(\hat{\boldsymbol{u}} - \hat{\boldsymbol{u}}_0), \quad \pi/2 \leq \beta_1 \leq \pi.
\end{equation}
At the bottom surface \( z=0 \), there is no light reflection:
\begin{equation}
G(x, y, 0, \beta_1, \beta_2) = 0, \quad 0 \leq \beta_1 \leq \pi/2.
\end{equation}
Here \( G_0 \) is the incident irradiation, \( \delta \) is the Dirac-delta function, and the incident light direction \( \hat{\boldsymbol{u}}_0 = -\hat{\boldsymbol{z}} \). By the definition of the delta function:
\begin{equation}
\int_0^{4\pi} f(\hat{\boldsymbol{u}}) \delta(\hat{\boldsymbol{u}} - \hat{\boldsymbol{u}}_0) d\omega = f(\hat{\boldsymbol{u}}_0).
\end{equation}

\subsection{Cell Swimming Velocity and Phototaxis}  

The average swimming velocity \( \boldsymbol{W}_c \) of microorganisms is given as
\begin{equation}
\boldsymbol{W}_c = W_s \boldsymbol{q},
\end{equation}
where \( W_s \) is the average cell swimming speed, and \( \boldsymbol{q} \) is the average cell swimming direction
\begin{equation}
\boldsymbol{q} = P(\mathbb{J}) \hat{\boldsymbol{z}}.
\end{equation}
The phototaxis function \( P(\mathbb{J}) \) depends on the total intensity \( \mathbb{J} \) and is defined as
\begin{equation}
P(\mathbb{J}) =
\begin{cases}
< 0 & \text{if } \mathbb{J}_c < \mathbb{J}, \\
\geq 0 & \text{if } \mathbb{J}_c \geq \mathbb{J}.
\end{cases}
\end{equation}
For specific microorganisms, \( P(\mathbb{J}) \) is expressed as\cite{ref9}
\begin{equation}
P(\mathbb{J}) = 0.8 \sin{(1.5\pi \varphi(\mathbb{J}))} - 0.1 \sin{(0.5\pi \varphi(\mathbb{J}))}, \quad
\varphi(\mathbb{J}) = \mathbb{J} \exp{(\xi(\mathbb{J} - 1))},
\end{equation}
where \( \xi \) is a parameter related to the critical intensity.

\subsection{Governing Equations}  

The continuity equation for an incompressible fluid is given by
\begin{equation}
\nabla \cdot \boldsymbol{v} = 0.
\end{equation}

Momentum Equation (under the Boussinesq approximation in a rotating porous medium):
\begin{equation}
\frac{\rho}{\phi} \left( \frac{\partial \boldsymbol{v}}{\partial t} + \frac{1}{\phi} \boldsymbol{v} \cdot \nabla \boldsymbol{v} + 2 \boldsymbol{\Omega} \times \boldsymbol{v} \right) = \mu \nabla^2 \boldsymbol{v} - \frac{\mu}{K} \boldsymbol{v} - \nabla \mathcal{P} - n \vartheta \Delta \rho g \hat{\boldsymbol{z}}.
\end{equation}

Here, \( \rho \) is the fluid density, \( \mu \) is the dynamic viscosity, \( \vartheta \) is the cell volume, \( \Delta \rho \) is the small density difference between cells and fluid, \( \phi \) is the porosity, \( K \) is permeability, and \( g \) is the gravitational acceleration.

Cell Conservation Equation:
\begin{equation}
\frac{\partial n}{\partial t} = -\nabla \cdot \boldsymbol{B}, \quad \boldsymbol{B} = n \boldsymbol{W}_c + n \boldsymbol{v} - \boldsymbol{D} \nabla n,
\end{equation}
where $\boldsymbol{D}$ is the diffusivity tensor.

\subsection{Boundary Conditions}  

At \( z=L \) (top boundary):
\begin{equation}
\boldsymbol{v} \cdot \hat{\boldsymbol{z}} = \frac{\partial^2}{\partial z^2} (\boldsymbol{v} \cdot \hat{\boldsymbol{z}}) = \boldsymbol{B} \cdot \hat{\boldsymbol{z}} = 0.
\end{equation}

At \( z=0 \) (bottom boundary):
\[
\boldsymbol{v} \cdot \hat{\boldsymbol{z}} = \boldsymbol{v} \times \hat{\boldsymbol{z}} = \boldsymbol{B} \cdot \hat{\boldsymbol{z}} = 0.
\]

The vertical velocity and cell flux vanish at both boundaries due to the conservation of mass and microorganisms. Depending on the nature of the boundary (rigid or stress-free), appropriate conditions are applied to the tangential velocity.

\subsection{Dimensionless Governing Equations} 
The governing equations of the bioconvection system are transformed into their dimensionless form by scaling all relevant physical quantities. Length, cell concentration, time, fluid velocity, and pressure are scaled by $L$, $n^\dag$, $\frac{L^2}{D}$, $\frac{D}{L}$, and $\frac{\mu D}{ L^2}$, respectively.

Substituting the non-dimensional variables into the original system of equations yields the following dimensionless form
\begin{equation}
\label{eqn:continuity_dimless}
    \boldsymbol{\nabla} \cdot \boldsymbol{v} = 0,
\end{equation}
\begin{equation}
\label{eqn:momentum_dimless}
    \frac{1}{S_c \phi} \left( \frac{\partial \boldsymbol{v}}{\partial t} + \frac{1}{\phi} \boldsymbol{v} \cdot \boldsymbol{\nabla} \boldsymbol{v} \right) + \frac{1}{\phi} T_a^{1/2} (\hat{\boldsymbol{z}} \times \boldsymbol{v}) = \nabla^2 \boldsymbol{v} - \frac{1}{D_a} \boldsymbol{v} - \boldsymbol{\nabla} \mathcal{P} - n R_a \hat{\boldsymbol{z}},
\end{equation}
\begin{equation}
\label{eqn:cell_conservation_dimless}
    \frac{\partial n}{\partial t} = -\boldsymbol{\nabla} \cdot \left( n W \boldsymbol{q} + n \boldsymbol{v} - \boldsymbol{\nabla} n \right).
\end{equation}

Here, the non-dimensional parameters are: \( S_c = \frac{\nu}{D} \) (Schmidt number),
\( T_a = \frac{4\Omega^2 L^4}{\nu^2} \) (Taylor number), \( W = \frac{W_s L}{D} \) (dimensionless swimming speed), \( R_a = \frac{n^\dag \vartheta \Delta \rho g L^3}{\mu D} \) (Rayleigh number), \( D_a = \frac{K}{L^2} \) (Darcy number), \( \nu = \frac{\mu}{\rho} \) (kinematic viscosity).
  
The non-dimensional boundary conditions for the system are given as follows:

At the Upper Boundary \( z = 1 \),
\begin{equation}
\label{eqn:upper_bc}
    \boldsymbol{v} \cdot \hat{\boldsymbol{z}} = \frac{\partial^2}{\partial z^2} (\boldsymbol{v} \cdot \hat{\boldsymbol{z}}) = \left( n W \boldsymbol{q} + n \boldsymbol{v} - \boldsymbol{\nabla} n \right) \cdot \hat{\boldsymbol{z}} = 0.
\end{equation}

At the Lower Boundary \( z = 0 \),
\begin{equation}
\label{eqn:lower_bc}
    \boldsymbol{v} \cdot \hat{\boldsymbol{z}}  = \boldsymbol{v} \times \hat{\boldsymbol{z}} = \left( n W \boldsymbol{q} + n \boldsymbol{v} - \boldsymbol{\nabla} n \right) \cdot \hat{\boldsymbol{z}} = 0.
\end{equation}

The non-dimensional radiative transfer equation for a purely absorbing medium (no scattering) is
\begin{equation}
\label{eqn:dimless_RTE}
    \frac{d G}{d u} + \chi n G = 0,
\end{equation}
where \( \chi = \iota n^\dag L \) represents the optical depth of the suspension.

The dimensionless boundary conditions for the radiative intensity \( G \) at the top and bottom boundaries became
\begin{equation}
\label{eqn:rte_bc_bottom}
    G(x, y, 0, \beta_1, \beta_2) = 0, \quad \text{for} \quad, 0 \leq \beta_1 \leq \frac{\pi}{2},
\end{equation}
\begin{equation}
\label{eqn:rte_bc_top}
    G(x, y, 1, \beta_1, \beta_2) = G_0 \delta(\hat{\boldsymbol{u}} - \hat{\boldsymbol{u}}_0), \quad \text{for} \quad, \frac{\pi}{2} \leq \beta_1 \leq \pi.
\end{equation}

\section{Basic State}
\label{Basic State}
In the basic state, the flow velocity \(\boldsymbol{v}\) is zero, the pressure is \(\mathcal{P} = \mathcal{P}_b\), the cell concentration is \(n = n_b(z)\), the radiative intensity is \(G = G_b(z, \beta_1)\), and the total intensity is \(\mathbb{J} = \mathbb{J}_b(z)\).

The steady-state form of the cell conservation equation becomes
\begin{equation}
\label{eqn:basic_cell_conservation}
\frac{d n_b}{dz} - W P_b n_b = 0,
\end{equation}
where \(P_b\) is the basic state phototaxis function. This equation is subject to the normalization condition
\begin{equation}
\label{eqn:basic_normalization}
\int_0^1 n_b \, dz = 1.
\end{equation}

The steady-state radiative transfer equation for the radiative intensity \(G_b\) is
\begin{equation}
\label{eqn:basic_RTE}
\frac{\partial G_b}{\partial z} + \frac{\chi n_b}{\cos{\beta_1}} G_b = 0,
\end{equation}
where \(\chi\) is the optical depth. The boundary condition at the top boundary (\(z = 1\)) is
\begin{equation}
\label{eqn:basic_RTE_BC}
G_b(1, \beta_1) = G_0 \delta(\hat{\boldsymbol{u}} - \hat{\boldsymbol{u}}_0).
\end{equation}

The solution of equation \eqref{eqn:basic_RTE} is
\begin{equation}
\label{eqn:basic_Gb}
G_b = G_0 \exp{\left( \frac{-\chi}{\cos{\beta_1}} \int_1^{z} n_b \, dz \right)} \delta(\hat{\boldsymbol{u}} - \hat{\boldsymbol{u}}_0).
\end{equation}

The total radiative intensity \(\mathbb{J}_b\) at steady state is obtained by integrating \(G_b\) over all solid angles (\(d\omega\)) as
\begin{eqnarray}
\label{eqn:basic_total_intensity}
\mathbb{J}_b = \int_0^{4\pi} G_b(z, \beta_1) \, d\omega 
= G_0 \exp{\left( \chi \int_1^{z} n_b \, dz \right)}.
\end{eqnarray}

The radiative heat flux \(\boldsymbol{f}_b\) at steady state is calculated as
\begin{eqnarray}
\label{eqn:basic_heat_flux}
\boldsymbol{f}_b = \int_0^{4\pi} G_b(z, \beta_1) \hat{\boldsymbol{u}} \, d\omega 
= -G_0 \exp{\left( \chi \int_1^{z} n_b \, dz \right)} \hat{\boldsymbol{z}} = |\boldsymbol{f}_b| (-\hat{\boldsymbol{z}}).
\end{eqnarray}

Here, the flux is directed along the negative \(z\)-direction. The average swimming orientation \(\boldsymbol{q}_b\) at the basic state is given by
\begin{equation}
\label{eqn:basic_swimming_orientation}
\boldsymbol{q}_b = -P_b \frac{\boldsymbol{f}_b}{|\boldsymbol{f}_b|} = P(\mathbb{J}_b) \hat{\boldsymbol{z}}.
\end{equation}

Here, \(P(\mathbb{J}_b)\) represents a phototaxis function of the total intensity \(\mathbb{J}_b\), and the swimming direction aligns with the \(z\)-axis.

\section{Linear Analysis}
\label{Linear Analysis}
To examine the linear instability of the basic state, a small perturbation \(\epsilon\) (\(0 < \epsilon \ll 1\)) is introduced into the basic state variables as
\[
\mathcal{P} = \mathcal{P}_b + \epsilon \bar{\mathcal{P}} + O(\epsilon^2), \quad n = n_b + \epsilon \bar{n}(x,y,z,t) + O(\epsilon^2), \quad \boldsymbol{v} = \boldsymbol{0} + \epsilon \bar{\boldsymbol{v}}(x,y,z,t) + O(\epsilon^2),
\]
where \(\bar{\boldsymbol{v}} = (v_1, v_2, v_3)\).

Substituting these perturbations into the governing equations, the linearized system is obtained
\begin{equation}
\label{eqn:equation28a}
\nabla \cdot \bar{\boldsymbol{v}} = 0,
\end{equation}

\begin{equation}
\label{eqn:equation28b}
\frac{1}{S_c \phi} \frac{\partial \bar{\boldsymbol{v}}}{\partial t} + \frac{1}{\phi} T_a^{1/2} (\hat{\boldsymbol{z}} \times \bar{\boldsymbol{v}}) = \nabla^2 \bar{\boldsymbol{v}} - \frac{1}{D_a} \bar{\boldsymbol{v}} - \boldsymbol{\nabla} \bar{\mathcal{P}} - \bar{n} R_a \hat{\boldsymbol{z}},
\end{equation}

\begin{equation}
\label{eqn:equation29}
\frac{\partial \bar{n}}{\partial t} = -\frac{d n_b}{dz} v_3 + \nabla^2 \bar{n} - W \boldsymbol{\nabla} \cdot (\bar{n} \boldsymbol{q}_b + n_b \bar{\boldsymbol{q}}).
\end{equation}

The total intensity \(\mathbb{J}\) can be expanded as
\[
\mathbb{J} = \mathbb{J}_b + \epsilon \bar{\mathbb{J}} + O(\epsilon^2) = G_0 \exp{\left( \chi \int_1^z (n_b + \epsilon \bar{n} + O(\epsilon^2)) \, dz \right)}.
\]
Collecting terms at \(O(\epsilon)\), we get
\begin{equation}
\label{eqn:linear_intensity}
\bar{\mathbb{J}} = G_0 \left( \chi \int_1^z \bar{n} \, dz \right) \exp{\left( \chi \int_1^z n_b \, dz \right)}.
\end{equation}

The average swimming orientation can be expressed as
\begin{equation}
\label{eqn:equation30}
\boldsymbol{q} = \boldsymbol{q}_b + \epsilon \bar{\boldsymbol{q}} + O(\epsilon^2) = P(\mathbb{J}_b + \epsilon \bar{\mathbb{J}} + O(\epsilon^2)) \hat{\boldsymbol{z}}.
\end{equation}
Collecting terms at \(O(\epsilon)\), we obtain
\begin{equation}
\label{eqn:equation31}
\bar{\boldsymbol{q}} = \bar{\mathbb{J}} \frac{\partial P}{\partial \mathbb{J}_b} \hat{\boldsymbol{z}}.
\end{equation}

Substituting \eqref{eqn:linear_intensity} and \eqref{eqn:equation31} into the linearized cell conservation equation \eqref{eqn:equation29}, we get
\begin{eqnarray}
\label{eqn:equation33}
\frac{\partial \bar{n}}{\partial t} - \nabla^2 \bar{n} - \chi W \frac{\partial}{\partial z} \left( n_b \mathbb{J}_b \frac{d P_b}{d \mathbb{J}_b} \right) \int_z^1 \bar{n} \, dz + 2 \chi n_b \mathbb{J}_b \frac{d P_b}{d \mathbb{J}_b} \bar{n} W + P_b \frac{\partial \bar{n}}{\partial z} W = -v_3 \frac{d n_b}{d z}.
\end{eqnarray}

Taking the vertical component of the curl applied twice to equation \eqref{eqn:equation28b}, we obtain the vertical velocity equation
\begin{equation}
\label{eqn:equation34}
\frac{1}{S_c \phi} \frac{\partial}{\partial t} (\nabla^2 v_3) + \frac{T_a^{1/2}}{\phi} \frac{\partial \zeta}{\partial z} = \nabla^4 v_3 - \frac{1}{D_a} \nabla^2 v_3 - R_a \nabla_H^2 \bar{n}.
\end{equation}

Here, \(\zeta\) is the vertical component of the vorticity.

Taking the curl of equation \eqref{eqn:equation28b}, we obtain the vorticity equation
\begin{equation}
\label{eqn:equation35}
\frac{1}{S_c \phi} \frac{\partial \zeta}{\partial t} - \frac{T_a^{1/2}}{\phi} \frac{\partial v_3}{\partial z} = \nabla^2 \zeta - \frac{1}{D_a} \zeta.
\end{equation}

The linearized boundary conditions at $z = 1$ are
\begin{equation*}
v_3 = \frac{\partial^2 v_3}{\partial z^2} = \frac{\partial \zeta}{\partial z} = W P_b \bar{n} - \frac{\partial \bar{n}}{\partial z} = 0.
\end{equation*}

At \(z = 0\), the boundary conditions are
\begin{equation*}
v_3 = \frac{\partial v_3}{\partial z} = \zeta = \frac{\partial \bar{n}}{\partial z} + W n_b \mathbb{J}_b \frac{d P_b}{d \mathbb{J}_b} \int_z^1 \bar{n} \, dz - W P_b \bar{n} = 0.
\end{equation*}

\section{Normal Mode Analysis}
\label{Normal Mode Analysis}
The perturbations $\bar{n}, v_3$, and $\zeta$ are decomposed into normal modes as follows:
\begin{eqnarray*}
    v_3 = V_3(z) \exp{[\eta t + i (\lambda_1 x + \lambda_2 y)]}, \\
    \quad \zeta = Z(z) \exp{[\eta t + i (\lambda_1 x + \lambda_2 y)]}, \\
    \quad \bar{n} = N_3(z) \exp{[\eta t + i (\lambda_1 x + \lambda_2 y)]},
\end{eqnarray*}
where $\lambda_1$ and $\lambda_2$ are the horizontal wavenumbers, and $\lambda = \sqrt{\lambda_1^2 + \lambda_2^2}$ is the resultant horizontal wavenumber.

Substituting the normal mode form into the linearized equations and simplifying, we obtain
\begin{eqnarray}
    \label{eqn:V3_eqn_new}
    \frac{\eta}{S_c \phi} \left( \frac{d^2}{dz^2} - \lambda^2 \right) V_3 - \left( \frac{d^2}{dz^2} - \lambda^2 \right)^2 V_3 + \frac{1}{D_a} \left( \frac{d^2}{dz^2} - \lambda^2 \right) V_3 = -\frac{T_a^{1/2}}{\phi} \frac{d Z}{d z} + \lambda^2 R_a N_3,
\end{eqnarray}
\begin{equation}
    \label{eqn:Z_eqn_new}
    \frac{\eta}{S_c \phi} Z = \frac{T_a^{1/2}}{\phi} \frac{d V_3}{d z} + \left( \frac{d^2}{dz^2} - \lambda^2 \right) Z - \frac{1}{D_a} Z,
\end{equation}
\begin{eqnarray}
    \label{eqn:N3_eqn_new}
    \chi W \frac{\partial}{\partial z} \left( n_b \mathbb{J}_b \frac{d P_b}{d \mathbb{J}_b} \right) \int_z^1 N_3 \, dz - \left( \eta + \lambda^2 + 2 \chi W n_b \mathbb{J}_b \frac{d P_b}{d \mathbb{J}_b} \right) N_3 
    - W P_b \frac{d N_3}{d z} + \frac{d^2 N_3}{d z^2} = \frac{d n_b}{d z} V_3.
\end{eqnarray}

\begin{table}
\caption{Typical parameters for the suspension of \textit{Chlamydomonas} phototactic microorganism \cite{rajput2024mathematical,kumar2023effect}.}
\begin{tabular}{ p{9cm} p{4cm}}
\hline
\hline

Kinematic viscosity &$\nu=10^{-2}$ cm$^2$/s\\
Cell volume &$\vartheta=5\times10^{-10}$ cm$^3$\\
Schmidt number &$S_c=20$ \\
Cell diffusivity &$D=5\times10^{-4}$ cm$^2$/s\\
Porosity of the medium &$\phi=0.76$\\
Scaled average swimming speed &$W=20L$\\
Cell radius &$a=10^{-3}$ cm\\
Ratio of cell density &$\Delta \varrho/\varrho=5\times 10^{-2}$ \\
Average concentration &$n^\dag=10^6$ cm$^{-3}$\\
Average cell swimming speed &$U_c =10^{-2}$ cm/s\\
\hline
\hline
\end{tabular}\\
\label{tab:table1}
\end{table}

Introducing the integral variable,
\begin{equation}
\label{eqn:N3_prime_new}
    N_3'(z) = \int_z^1 N_3 \, dz,
\end{equation}
the equations are rewritten as
\begin{eqnarray}
    \label{eqn:V3_final}
    \frac{d^4 V_3}{d z^4} - \left( 2\lambda^2 + \frac{\eta}{S_c \phi} + \frac{1}{D_a} \right) \frac{d^2 V_3}{d z^2} + \lambda^2 \left( \lambda^2 + \frac{\eta}{S_c \phi} + \frac{1}{D_a} \right) V_3 = \frac{T_a^{1/2}}{\phi} \frac{d Z}{d z} + \lambda^2 R_a \frac{d N_3'}{d z},
\end{eqnarray}
\begin{equation}
    \label{eqn:Z_final}
    \frac{d^2 Z}{d z^2} - \left( \lambda^2 + \frac{\eta}{S_c \phi} + \frac{1}{D_a} \right) Z = -\frac{T_a^{1/2}}{\phi} \frac{d V_3}{d z},
\end{equation}
\begin{eqnarray}
    \label{eqn:N3_final}
    \frac{d^3 N_3'}{d z^3} - W P_b \frac{d^2 N_3'}{d z^2} - \left( \eta + \lambda^2 + 2\chi W n_b \mathbb{J}_b \frac{d P_b}{d \mathbb{J}_b} \right) \frac{d N_3'}{d z} 
    - \chi W \frac{d}{d z} \left( n_b \mathbb{J}_b \frac{d P_b}{d \mathbb{J}_b} \right) N_3' = -\frac{d n_b}{d z} V_3.
\end{eqnarray}

The corresponding boundary conditions become
\begin{equation*}
    V_3 = \frac{d^2 V_3}{d z^2} = \frac{d Z}{d z} = W P_b \frac{d N_3'}{d z} - \frac{d^2 N_3'}{d z^2}= 0 \quad \text{at} \quad z = 1,
\end{equation*}
\begin{equation*}
    V_3 = \frac{d V_3}{d z} = Z = \chi W n_b \mathbb{J}_b \frac{d P_b}{d \mathbb{J}_b} N_3' + W P_b \frac{d N_3'}{d z} - \frac{d^2 N_3'}{d z^2} = 0 \quad \text{at} \quad z = 0,
\end{equation*}
\begin{equation*}
    N_3'(z) = 0 \quad \text{at} \quad z = 1.
\end{equation*}

The presence of rotational effects, quantified by Taylor number $T_a$, modifies the stability behavior compared to the non-rotating case. When $T_a = 0$, the above system reduces to the model analyzed in \citet{rajput2024mathematical}. In the case of a non-porous medium, the above system simplifies to the rotating phototactic bioconvection model previously analyzed in \citet{kumar2023effect}.

\begin{figure*}[t]
    \centering
    \includegraphics[width=16cm, height=11cm ]{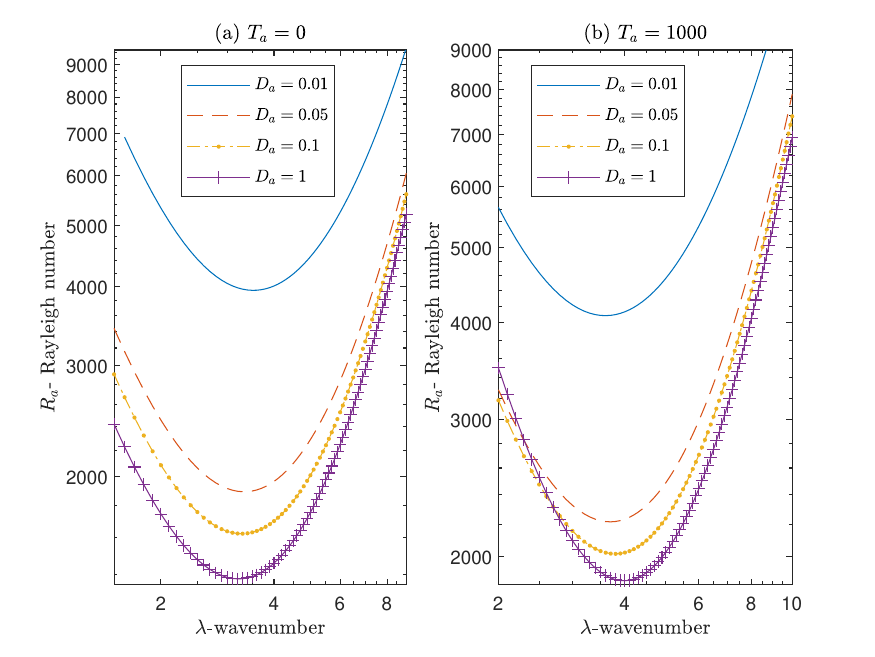}
    \caption{Neutral curves plotted against the wavenumber for the fixed parameters  $W=10$, $\chi=0.5$, $\mathbb{J}_c=0.63$, (a) $T_a=0$, (b) and $T_a=1000$ as the Darcy number $D_a$ varies. }
   \label{fig:10v0.5k0.63Ic.pdf}
 \end{figure*}

\begin{table}[t]
\caption{\label{tab:table2} For $W=10$, $\chi=0.5$, $\mathbb{J}_c=0.63$, $\xi=-0.068$, the critical wavenumber and the Rayleigh number.}
\begin{ruledtabular}
\begin{tabular}{ccccc}
$T_a$& $D_a$ & $\lambda^c$&$R_a^c$&$Im(\eta)$ \\
\hline
0    & 1   & 3.2 & 1377.22 & 0 \\ 
0    & 0.5 & 3.2 & 1405.27 & 0 \\ 
0    & 0.1 & 3.3 & 1624.41 & 0 \\ 
0    & 0.05 & 3.3 & 1893.42 & 0 \\ 
0    & 0.01 & 3.5 & 3950.54 & 0 \\ 
100  & 1   & 3.3 & 1439.00 & 0 \\ 
100  & 0.5 & 3.3 & 1464.15 & 0 \\ 
100  & 0.1 & 3.3 & 1669.51 & 0 \\ 
100  & 0.05 & 3.4 & 1928.43 & 0 \\ 
100  & 0.01 & 3.5 & 3964.27 & 0 \\ 
1000 & 1   & 4.0 & 1860.05 & 0 \\ 
1000 & 0.5 & 4.0 & 1875.73 & 0 \\ 
1000 & 0.1 & 3.8 & 2014.62 & 0 \\ 
1000 & 0.05 & 3.7 & 2215.68 & 0 \\ 
1000 & 0.01 & 3.6 & 4088.21 & 0 \\ 
\end{tabular}
\end{ruledtabular}
\end{table}

\begin{figure}[t]
    \centering
    \includegraphics[width=16cm, height=11cm ]{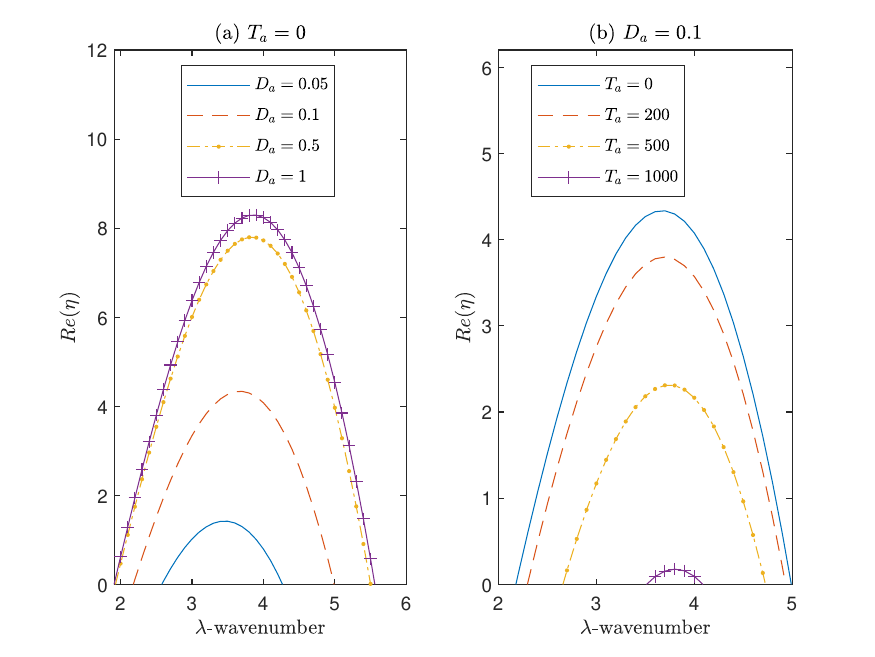}
    \caption{Growth rate curves are plotted against the wavenumber for the following fixed parameters $W=10$, $\chi=0.5$, $\mathbb{J}_c=0.63$, $R_a=2030$. (a) Displays growth rate curves for $T_a=0$ as the Darcy number $D_a$ varies. (b) Displays growth rate curves for $D_a=0.1$ as the Taylor number $T_a$ varies.}
   \label{fig:growth.pdf}
 \end{figure}

\begin{figure*}[t]
    \centering
    \includegraphics[width=16cm, height=11cm ]{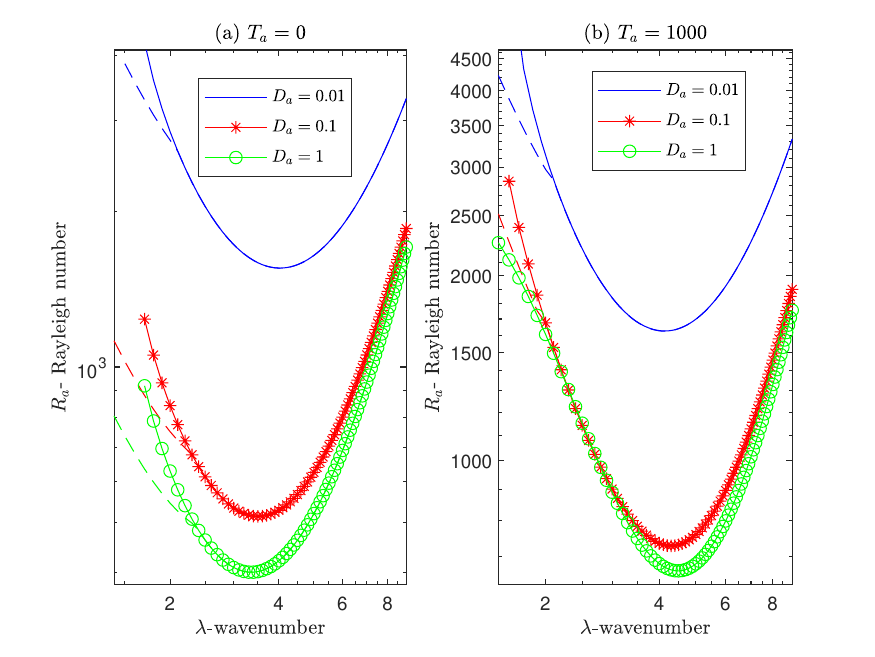}
    \caption{Neutral curves plotted against the wavenumber for the fixed parameters  $W=15$, $\chi=0.5$, $\mathbb{J}_c=0.68$, (a) $T_a=0$, (b) and $T_a=1000$ as the Darcy number $D_a$ varies. The oscillating branches are represented by dotted lines.}
   \label{fig:15v0.5k0.68Gc.pdf}
 \end{figure*}

\begin{table}
\caption{\label{tab:table3} For $W=15$, $\chi=0.5$, $\mathbb{J}_c=0.68$, $\xi=0.165$, the critical wavenumber and the Rayleigh number. A $*$ symbol denotes that the neutral curve includes an oscillatory branch.}
\begin{ruledtabular}
\begin{tabular}{ccccc}
$T_a$& $D_a$ & $\lambda^c$&$R_a^c$&$Im(\eta)$ \\
\hline
0    & 1   & 3.4 & $401.21 ^*$  & 0 \\ 
0    & 0.5 & 3.4 & $414.09 ^*$ & 0 \\ 
0    & 0.1 & 3.5 & $514.73 ^*$ & 0 \\ 
0    & 0.01 & 4.0 & $1552.85^*$ & 0 \\ 
100  & 1   & 3.6 & $437.59 ^*$ & 0 \\ 
100  & 0.5 & 3.6 & $448.85 ^* $& 0 \\ 
100  & 0.1 & 3.6 & $540.48^* $ & 0 \\ 
100  & 0.01 & 4.0 & $1560.04^*$ & 0 \\ 
1000 & 1   & 4.5 & 662.67  & 0 \\ 
1000 & 0.5 & 4.5 & 669.28  & 0 \\ 
1000 & 0.1 & 4.3 & $727.78^* $ & 0 \\ 
1000 & 0.01 & 4.1 & $1625.53^*$ & 0 \\  
\end{tabular}
\end{ruledtabular}
\end{table}

\begin{figure*}[t]
    \centering
    \includegraphics[width=16cm, height=11cm ]{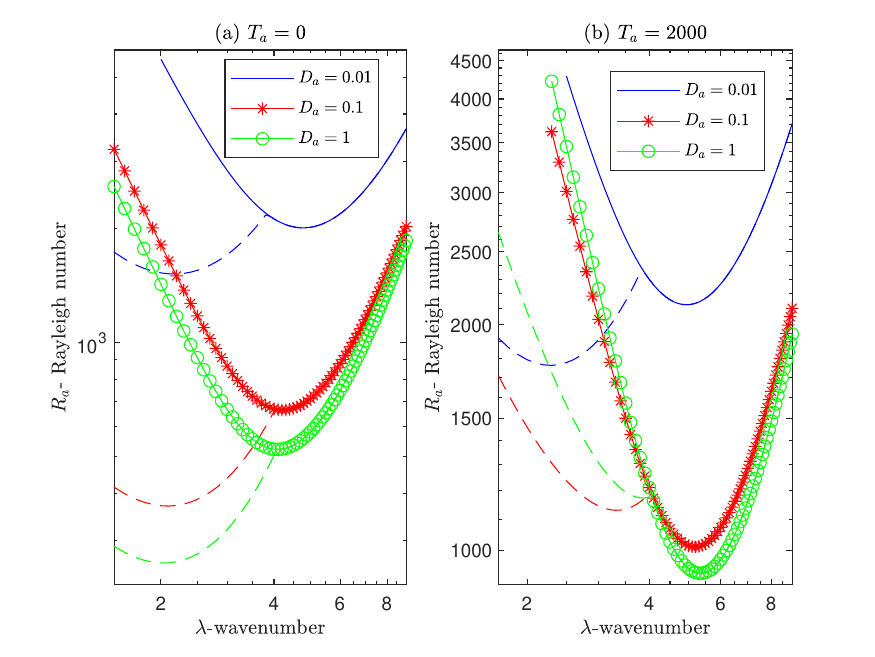}
    \caption{Neutral curves plotted against the wavenumber for the fixed parameters  $W=15$, $\chi=1.0$, $\mathbb{J}_c=0.51$, (a) $T_a=0$, (b) and $T_a=2000$ as the Darcy number $D_a$ varies. The oscillating branches are represented by dotted lines. }
   \label{fig:15v1.0k0.51Gc.pdf}
 \end{figure*}

\begin{figure*}[t]
    \centering
    \includegraphics[width=16cm, height=11cm ]{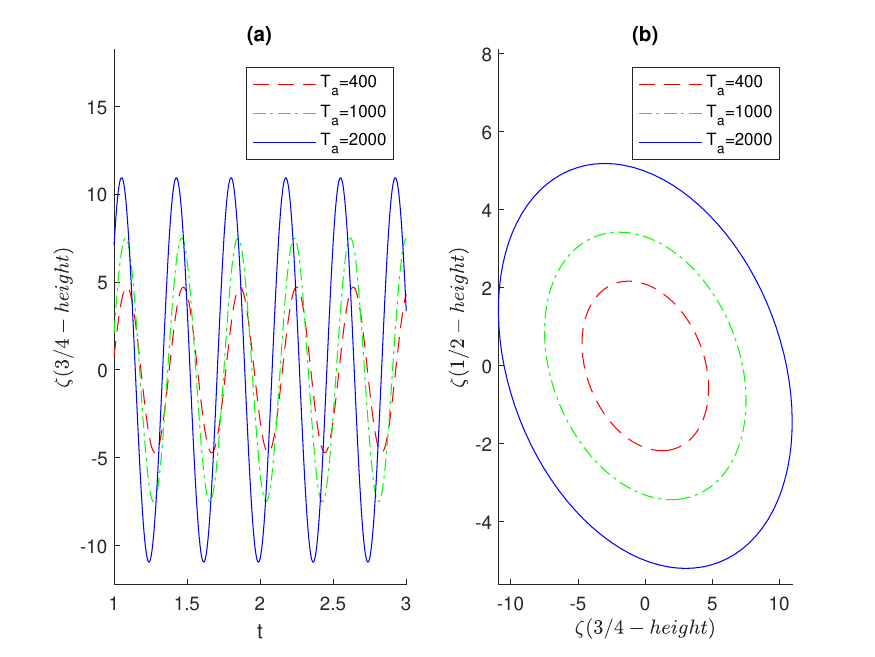}

    \caption{(a) Time-evolving perturbed fluid vorticity \(\zeta \) and (b) phase diagram are shown for the fixed parameters \( W = 15 \), \( \chi = 1.0 \), \( \mathbb{J}_c=0.51 \), and $D_a=0.01$.}
   \label{fig:limitcycle.pdf}
 \end{figure*}

\begin{table}
\caption{\label{tab:table4} For $W=15$, $\chi=1.0$, $\mathbb{J}_c=0.51$, $\xi=-0.485$, the critical wavenumber and the Rayleigh number. A result marked with a $\#$ symbol signifies that the critical values are located on an oscillatory branch, while a $*$ symbol denotes that the neutral curve includes an oscillatory branch.}
\begin{ruledtabular}
\begin{tabular}{ccccc}
$T_a$& $D_a$ & $\lambda^c$&$R_a^c$&$Im(\eta)$ \\
\hline
0    & 1.0 & 2.0 & $261.87^{\#} $ & 12.13 \\ 
0    & 0.5 & 2.0 & $273.87^{\#}$  & 12.21 \\ 
0    & 0.1 & 2.1 & $370.01^{\#}$  & 13.06 \\ 
0    & 0.01 & 2.2 & $1514.10^{\#}$ & 16.14 \\ 
400  & 1.0 & 2.8 & $500.78^{\#} $ & 15.34 \\ 
400  & 0.5 & 2.7 & $503.73^{\#}$  & 15.28 \\ 
400  & 0.1 & 2.5 & $546.87^{\#}$  & 15.06 \\ 
400  & 0.01 & 2.2 & $1564.08^{\#}$ & 16.21 \\ 
2000 & 1.0 & 5.3 & $931.34^* $ & 0.00 \\ 
2000 & 0.5 & 5.3 & $939.30^* $ & 0.00 \\ 
2000 & 0.1 & 5.2 & $1009.80^*$ & 0.00 \\ 
2000 & 0.01 & 2.3 &$ 1764.13^{\#}$ & 16.77 \\   
\end{tabular}
\end{ruledtabular}
\end{table}

  \begin{figure*}[t]
    \centering
    \includegraphics[width=16cm, height=11cm ]{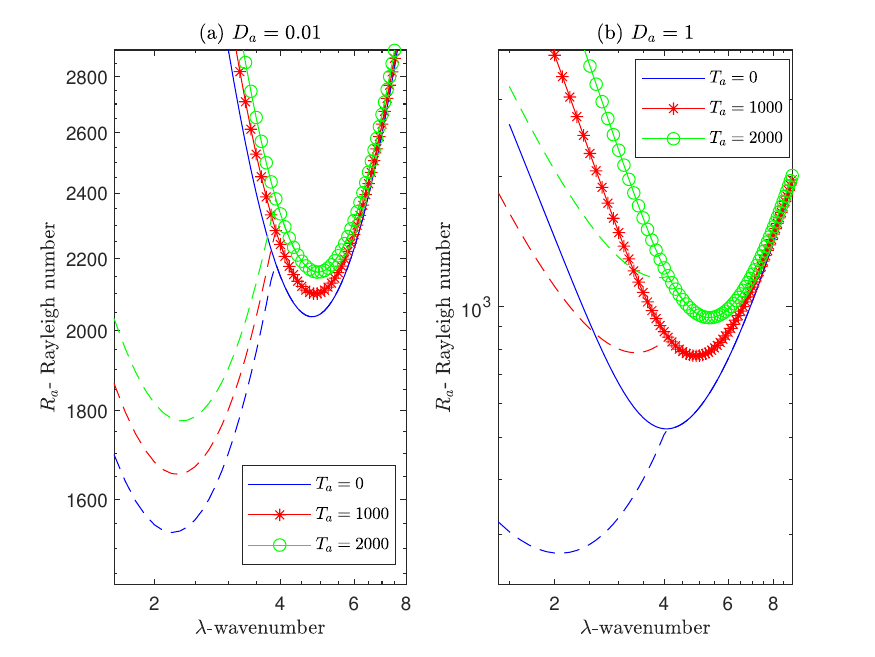}
    \caption{Neutral curves plotted against the wavenumber for the fixed parameters  $W=20$, $\chi=0.5$, $\mathbb{J}_c=0.64$, (a) $D_a=0.01$, (b) and $D_a=1$ as the Taylor number $T_a$ varies. The oscillating branches are represented by dotted lines.}
   \label{fig:fig20v0.5k0.64Gc.pdf}
 \end{figure*}

\begin{table}[t]
\caption{\label{tab:table5} For $W=20$, $\chi=0.5$, $\mathbb{J}_c=0.64$, $\xi=-0.0275$, the critical wavenumber and the Rayleigh number. A result marked with a $\#$ symbol signifies that the critical values are located on an oscillatory branch, while a $*$ symbol denotes that the neutral curve includes an oscillatory branch.}
\begin{ruledtabular}
\begin{tabular}{ccccc}
$T_a$& $D_a$ & $\lambda^c$&$R_a^c$&$Im(\eta)$ \\
\hline
0    & 1.0 & 2.1 & $273.33^{\#}$  & 12.09 \\ 
100  & 1.0 & 2.3 & $339.17^{\#}$  & 13.08 \\ 
1000 & 1.0 & 4.9 & $771.44^*$  & 0.00 \\ 
2000 & 1.0 & 5.3 & $944.83^*$  & 0.00 \\ 
0    & 0.1 & 2.1 & $381.44^{\#}$  & 12.73 \\ 
100  & 0.1 & 2.2 & $427.66^{\#}$  & 13.27 \\ 
1000 & 0.1 & 2.9 & $788.35^{\#}$  & 15.86 \\ 
2000 & 0.1 & 5.2 & $1024.38^*$ & 0.00 \\ 
0    & 0.01 & 2.2 & $1531.56^{\#}$ & 15.59 \\ 
100  & 0.01 & 2.2 & $1543.70^{\#}$ & 15.61 \\ 
1000 & 0.01 & 2.3 & $1654.38^{\#}$ & 16.01 \\ 
2000 & 0.01 & 2.3 & $1775.36^{\#}$ & 16.18 \\   
\end{tabular}
\end{ruledtabular}
\end{table}

\begin{table}[t]
\caption{\label{tab:table6} For $W=20$, $\chi=1.0$, $\mathbb{J}_c=0.5$, $\xi=-0.5045$, the critical wavenumber and the Rayleigh number. A result marked with a $\#$ symbol signifies that the critical values are located on an oscillatory branch, while a $*$ symbol denotes that the neutral curve includes an oscillatory branch.}
\begin{ruledtabular}
\begin{tabular}{ccccc}
$T_a$& $D_a$ & $\lambda^c$&$R_a^c$&$Im(\eta)$ \\
\hline
0    & 1.0 & 1.9  & $186.7^{\#}$  & 14.77 \\ 
100  & 1.0 & 2.1  & $245.09^{\#}$ & 16.68 \\ 
1000 & 1.0 & 3.2  & $650.05^{\#}$ & 24.73 \\ 
2000 & 1.0 & 5.7  & $960.74^*$ & 0.00 \\ 
0    & 0.1 & 1.9  & $268.69^{\#}$ & 15.71 \\ 
100  & 0.1 & 2.0  & $307.41^{\#}$ & 16.66 \\ 
1000 & 0.1 & 2.7  & $630.45^{\#}$ & 23.01 \\ 
2000 & 0.1 & 3.2  & $954.00^{\#}$ & 26.14 \\ 
0    & 0.01 & 1.9 & $1158.82^{\#}$ & 19.73 \\ 
100  & 0.01 & 2.0 & $1165.84^{\#}$ & 20.27 \\ 
1000 & 0.01 & 2.0 & $1262.32^{\#} $& 20.62 \\ 
2000 & 0.01 & 2.1 & $1371.18^{\#} $& 21.57 \\    
\end{tabular}
\end{ruledtabular}
\end{table}

 \begin{figure*}[t]
    \centering
    \includegraphics[width=16cm, height=11cm ]{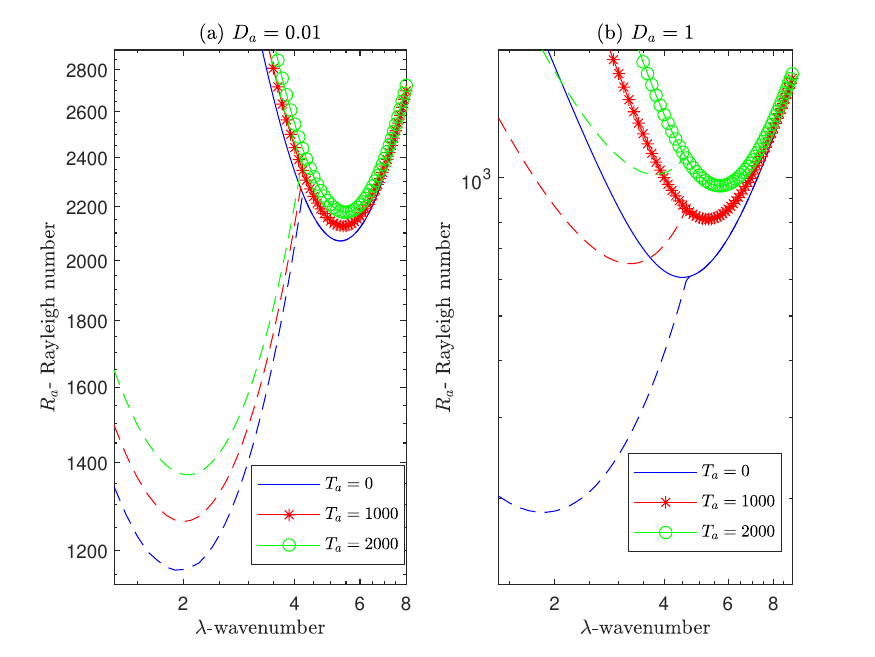}
    \caption{Neutral curves plotted against the wavenumber for the fixed parameters  $W=20$, $\chi=1.0$, $\mathbb{J}_c=0.5$, (a) $D_a=0.01$, (b) and $D_a=1$ as the Taylor number $T_a$ varies. The oscillating branches are represented by dotted lines.}
   \label{fig:fig20v1.0k0.51Gc.pdf}
 \end{figure*}


\section{Numerical solution}
\label{Numerical solution}
The system of ordinary differential equations (ODEs) \((\ref{eqn:V3_final})-(\ref{eqn:N3_final})\) forms a ninth-order system with corresponding nine boundary conditions. To solve this high-order system, we employ the MATLAB Bvp4c \cite{shampine2003solving}. Linear stability of the basic state is examined by plotting neutral curves in the \((\lambda, R_a)\) plane, where \(\lambda\) represents the wavenumber and \(R_a\) is the Rayleigh number. Neutral curves consist of points where the real part of eigenvalue \(\eta\) (growth rate) becomes zero.  
The nature of the perturbations along the neutral curve is determined based on the imaginary part of the eigenvalue, that is, \(Im(\eta)\). If \(Im(\eta) \neq 0\), the perturbations are oscillatory, indicating the presence of overstable solutions. In contrast, when \(Im(\eta) = 0\), the perturbations are stationary, and the principle of exchange of stabilities applies \citep{ref12}. The system is considered linearly stable when all perturbations exhibit a negative growth rate (\(Re(\eta) < 0\)) for all wavenumbers. However, as the Rayleigh number increases, a narrow band of wavenumbers may exhibit positive growth rates (\(Re(\eta) > 0\)), signifying the onset of instability. Of particular interest is the solution branch where the Rayleigh number \(R_a\) attains its minimum value \(R_a^c\). The corresponding wavenumber \(\lambda^c\) and Rayleigh number \(R_a^c\) represent the most unstable mode of bioconvection. To determine the model parameters, we assume phototactic microorganisms similar to \textit{Chlamydomonas}. To ensure consistency with previous studies on phototactic bioconvection, parameter values from \citet{rajput2024mathematical}, \citet{kumar2023effect} are adopted (see Table~\ref{tab:table1}). To facilitate comparison with existing phototactic bioconvection models, the Schmidt number \(S_c\) and incident light intensity \(G_0\) are fixed at \(20\) and \(0.8\), respectively. Furthermore, the critical intensity \(\mathbb{J}_c\) is related to the parameter \(\xi\) such that \(0.3 \leq \mathbb{J}_c \leq 0.8\) when \(-1.1 \leq \xi \leq 1.1\) \citep{ref17}.

\subsection{Effect of Darcy number}

Figure \ref{fig:10v0.5k0.63Ic.pdf} illustrates the neutral curves as a function of the wavenumber ($\lambda$) for a cell swimming speed of $W = 10$, absorption coefficient $\chi = 0.5$, critical total intensity $\mathbb{J}_c = 0.63$, (a) Taylor number $T_a=0$ and (b) $T_a=1000$. The curves demonstrate the impact of Darcy number $D_a$ on the stability threshold. For Taylor number $T_a = 0$, an increase in the Darcy number results in a decrease in the critical Rayleigh number, accompanied by an increase in the pattern wavelength, indicating enhanced instability in the system. The expected wavelengths of initial disturbances can be estimated by analyzing growth rate curves for Rayleigh numbers above the critical value. Figure \ref{fig:growth.pdf}(a) shows growth rate curves for $R_a = 2030$ and Darcy numbers $D_a = 0.05$, $0.1$, $0.5$, and $1$. For $D_a = 0.05$, the maximum growth rate occurs at $\lambda = 3.5$, corresponding to a pattern wavelength of $1.79$. As $D_a$ increases, the pattern wavelength decreases further. When $D_a = 1$, the maximum growth rate corresponds to a more realistic initial pattern wavelength of $1.61$. This trend highlights how increasing the Darcy number influences the system's initial instability patterns. A similar trend is observed for Taylor number $T_a = 100$, where the pattern wavelength increases as the Darcy number $D_a$ rises (refer to Table \ref{tab:table2}). However, when Taylor number $T_a$ is raised to $1000$, the behavior shifts; as the Darcy number $D_a$ increases, the pattern wavelength decreases. For $D_a = 0.1$, the variation of growth rate with increasing Taylor number $T_a$ from $0$ to $1000$ is shown in Figure 3(b). At $T_a = 0$, the maximum growth rate is $4.33$, occurring at a wavenumber $\lambda = 3.7$, corresponding to a pattern wavelength of $1.69$. As $T_a$ increases to $1000$, the maximum growth rate decreases to $0.17$, and the pattern wavelength reduces slightly to $1.65$. For $T_a = 1000$, the critical Rayleigh number becomes $R_a^c = 2014.62$, which approaches the fixed Rayleigh number $R_a = 2030$. Further increasing $T_a$ causes the corresponding critical Rayleigh number to surpass the fixed Rayleigh number, resulting in a negative growth rate. This shift pushes the system into the stable region, indicating a stabilization effect due to the increased Taylor number.

Figure \ref{fig:15v0.5k0.68Gc.pdf} illustrates the neutral curves plotted against the wavenumber for the fixed parameters \( W = 15 \), \( \chi = 0.5 \), and \( \mathbb{J}_c = 0.68 \): (a) for \( T_a = 0 \) as the Darcy number \( D_a \) varies, and (b) for \( T_a = 1000 \) as the Darcy number \( D_a \) varies. For the critical total intensity \( \mathbb{J}_c = 0.68 \), the basic state exhibits a maximum concentration at \( z = 0.9 \), where the sublayer forms. When \( T_a = 0 \) and \( D_a = 0.01 \), the critical Rayleigh number \( R_a = 401.21 \) is achieved at wavenumber \( \lambda = 3.4 \). The oscillatory branch persists up to \( \lambda = 2.4 \) but vanishes for higher wavenumbers. As the Darcy number \( D_a \) increases to $1$, similar behavior is observed with a reduction in the critical Rayleigh number, indicating enhanced instability. When the Taylor number \( T_a \) is increased to $100$, the oscillatory branch persists for all values of the Darcy number \( D_a \) (see Table \ref{tab:table3}). However, when \( T_a \) is raised further to $1000$, the oscillatory branch is observed only for \( D_a = 0.01 \) and \( D_a = 0.1 \). Upon increasing \( D_a \) to $0.5$ and $1$, the oscillatory branch disappears, leaving only the stationary branch.

Figure \ref{fig:15v1.0k0.51Gc.pdf} illustrates the neutral curves plotted against the wavenumber for the fixed parameters \( W = 15 \), \( \chi = 1.0 \), \( \mathbb{J}_c = 0.51 \), (a) \( T_a = 0 \), and (b) \( T_a = 2000 \), as the Darcy number \( D_a \) varies. For the critical total intensity \( \mathbb{J}_c = 0.51 \), the basic state exhibits maximum concentration at \( z = 0.75 \), where a sublayer forms. At \( T_a = 0 \) and \( D_a = 0.01 \), the critical Rayleigh number \( R_a^c = 261.87 \) is obtained at a wavenumber \( \lambda^c = 2.0 \). This corresponds to the most unstable solution on the oscillatory branch, with a positive frequency \( \text{Im($\eta$)} = 12.13 \). The time for one complete oscillation is \( {2\pi}/{\text{Im($\eta$)}} =0.51\) units. As the Darcy number increases up to \( D_a = 1 \), the most unstable solution remains oscillatory in nature. When the Taylor number \( T_a \) is increased to $400$, the most unstable solution remains on the oscillatory branch for all Darcy number values \( D_a \) from $0.01$ to $1$ (see Table \ref{tab:table4}). However, as the Taylor number further increases to $2000$, the behavior changes. For \( D_a = 0.01 \), the most unstable solution continues to reside on the oscillatory branch. For \( D_a = 0.1 \), $0.5$, and $1$, while the oscillatory branch still exists, the most unstable solution shifts to the stationary branch. Here, a transition from oscillatory solutions to stationary solutions is observed as the Darcy number increases. Figure \ref{fig:limitcycle.pdf} illustrates the predicted time-evolution of the perturbed fluid vorticity component \( \zeta \) and its corresponding phase diagram at critical wavenumbers for varying Taylor numbers \( T_a = 400, 1000, \) and \( 2000 \). Notably, the oscillation period, given by \( 2\pi/\text{Im($\eta$)}  \), serves as the bifurcation (control) parameter. Consequently, the destabilization of bioconvective flow leads to the emergence of a limit cycle characterized by an isolated trajectory. This transition, identified through bifurcation analysis, is recognized as a Hopf bifurcation triggered by flow instability. The results demonstrate that as Taylor number \( T_a \) increases, the amplitude of the oscillations in vorticity \( \zeta \) also increases, indicating stronger oscillations. 

\subsection{Effect of Rotation}

Figure \ref{fig:fig20v0.5k0.64Gc.pdf} presents the neutral curves plotted against the wavenumber for the fixed parameters \( W = 20 \), \( \chi = 0.5 \), and \( \mathbb{J}_c = 0.64 \) under two conditions: (a) \( D_a = 0 .01\) and (b) \( D_a = 1 \), as the Taylor number \( T_a \) varies. For the critical total intensity \( \mathbb{J}_c = 0.64 \), the maximum concentration in the basic state is observed at \( z = 0.75 \), where a sublayer forms. When \( D_a = 0 \) and \( T_a = 0 \), an oscillatory branch bifurcates from the stationary branch, with the most unstable solution being oscillatory. As the Taylor number increases to $2000$, oscillatory solutions persist, and the critical Rayleigh number rises, enhancing the system's stability. Stabilization occurs as rotation limits vertical motion, confining the fluid's movement primarily to the horizontal plane, thereby reducing the onset of bioconvection. Uniform rotation induces a Coriolis force, which creates a vortex that influences the fluid's flow. As the angular velocity increases, the fluid aligns more closely with the vortex lines. This alignment restricts perpendicular motion, making it challenging for convection-related streamlines to close \cite{chand1953}. Consequently, uniform rotation tends to enhance the system's stability. For \( D_a = 0.1 \), the results are summarized in Table \ref{tab:table5}. When \( D_a = 1 \), oscillatory solutions persist for \( T_a = 0 \) and \( T_a = 100 \). However, for \( T_a = 1000 \) and \( T_a = 2000 \), while the oscillatory branch continues to exist, the most unstable solution transitions to the stationary branch. This highlights the transition from oscillatory to stationary solutions as the Taylor number increases.

\begin{figure}[H]
    \centering
    \includegraphics[width=16cm, height=18cm ]{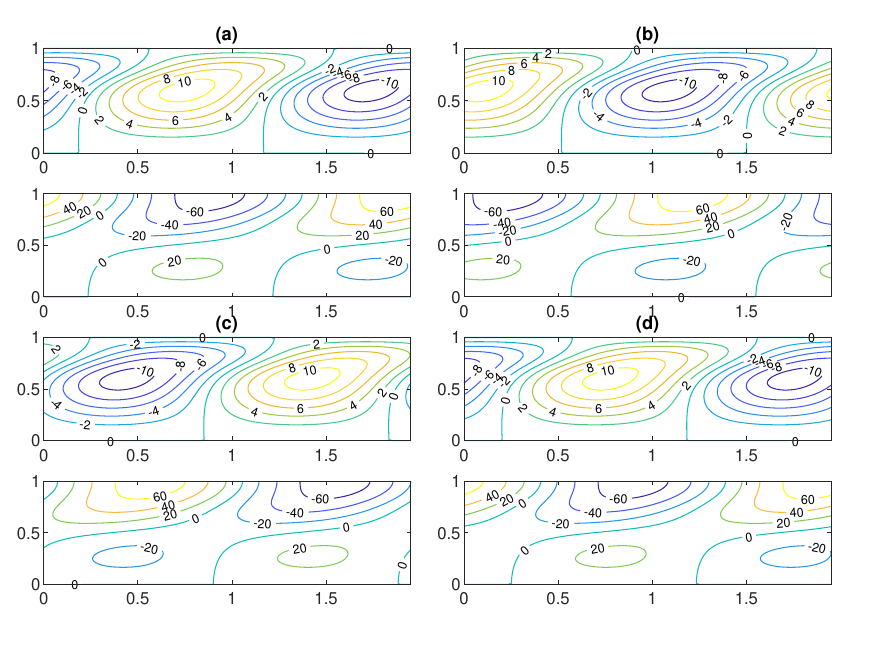}
    \caption{Flow pattern produced by the perturbed velocity $v_3$ (upper) and vorticity $\zeta$ (lower) throughout one cycle of oscillation for $W=20$, $\chi=1.0$, $\mathbb{J}_c=0.5$, $D_a=1$, $T_a=1000$, (a) time $t=0$, (b) $t=0.08$, (c) $t=0.16$, (d) $t=0.25$.}
   \label{fig:periodic.pdf}
 \end{figure}

Figure \ref{fig:fig20v1.0k0.51Gc.pdf} illustrates neutral curves plotted against the wavenumber for the fixed parameters $W=20$, $\chi=1.0$, and $\mathbb{J}_c=0.5$, with (a) $D_a=0.01$ and (b) $D_a=1$, as the Taylor number $T_a$ varies. For the critical total intensity $\mathbb{J}_c=0.5$, the basic state exhibits maximum concentration at \( z=0.75 \), forming a sublayer at this position. In this configuration, for \( D_a=0 \) and \( D_a=0.1 \), changes in \( T_a \) from $0$ to $2000$ result in only oscillatory solutions as the most unstable outcomes. However, when \( D_a=1 \), oscillatory solutions persist for \( T_a \) values ranging from $0$ to $1000$. For $T_a=1000$, the onset of overstability occurs at \( \lambda^c = 3.2 \) and \( R_a^c =650.05 \), where two complex conjugate eigenvalues (\( \eta = 0\pm 24.73 \)) emerge. This transition is classified as a Hopf bifurcation. The bioconvective flow patterns associated with these complex conjugate eigenvalues form mirror images of each other. The oscillation period is calculated as \( 2\pi/\text{Im($\eta$)}=0.25 \) units. The bioconvective fluid motions reach full nonlinearity on a timescale significantly shorter than the predicted period of overstability. Consequently, the perturbed eigenmodes (\( v_3 \) and \( \zeta \)) effectively reveal the flow patterns and fluid vorticity throughout a single oscillation cycle (refer to Fig. \ref{fig:periodic.pdf}). This analysis suggests the presence of a traveling wave solution propagating toward the left in the figure. Upon increasing \( T_a \) to $2000$, the oscillatory branch continues to exist, but the most unstable solutions transition to the stationary branch.

\begin{table}[H]
\caption{\label{tab:table7} Comparing the critical wavenumber and the critical Rayleigh number between the proposed model and a similar model in a non-rotating medium ($T_a=0$). A result marked with a $\#$ symbol signifies that the critical values are located on an oscillatory branch.}
\begin{ruledtabular}
\begin{tabular}{cccccccc}
$W$& $\chi$ & $\mathbb{J}_c$&$D_a$ &\multicolumn{2}{c}{Non-rotating model \cite{rajput2024mathematical}} & \multicolumn{2}{c}{Proposed model} \\ 
\cline{5-6} \cline{7-8}
&&&&$R_a^c$ &$\lambda^c$&$R_a^c$ &$\lambda^c$\\
\hline
10    & 0.5 & 0.63  &1 & 1377.21&3.2&1377.22&3.2 \\ 
10  & 0.5 & 0.63 & 0.1 & 1624.21&3.28&1624.41&3.3 \\ 
10 & 0.5 & 0.63  & 0.01 & 3949.35&3.52&3950.54&3.5 \\ 
15    & 1.0 & 0.51  &1 &$ 263.08^{\#}$&2.15&$261.87^{\#}$&3.2 \\ 
15  & 1.0 & 0.51 & 0.1 & $371.45^{\#}$&2.07&$370.01^{\#}$&3.3 \\ 
15 & 1.0 & 0.51  & 0.01 & $1519.83^{\#}$&2.01&$1514.1^{\#}$&2.2 \\ 
20 &0.5 & 0.64  & 1 & $271.17^{\#}$&2.07&$273.33^{\#}$&2.1 \\ 
20   & 0.5 & 0.64 & 0.1 & $381.17^{\#}$&2.07&$273.33^{\#}$&2.1 \\ 
20  & 0.5 & 0.64  & 0.01 &$ 1530.77^{\#}$&2.2&$1531.56^{\#}$&2.2 \\ 
20 &1.0 & 0.5  & 1 & $186.68^{\#}$&1.82&$186.7^{\#}$&1.9 \\ 
20   & 1.0 & 0.5 & 0.1 &$ 268.15^{\#}$&1.86&$268.69^{\#}$&1.9 \\ 
20  & 1.0 & 0.5  & 0.01 & $1151.25^{\#}$&1.95&$1158.82^{\#}$&1.9 \\    
\end{tabular}
\end{ruledtabular}
\end{table}

\subsection{Effect of critical total intensity}

\begin{figure*}
    \centering
    \includegraphics[width=16cm, height=11cm ]{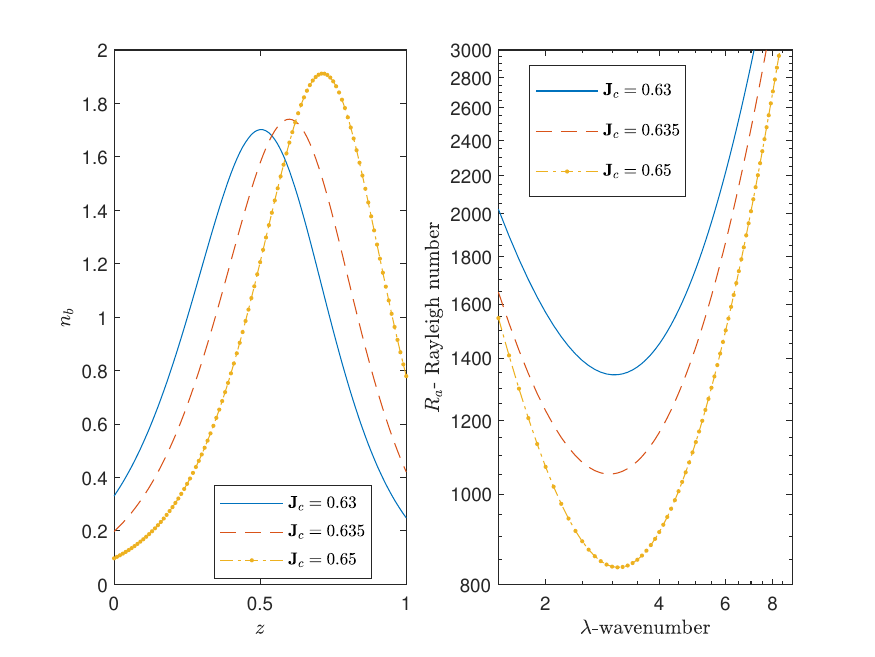}
    \caption{(a) Basic state concentration and (b) neutral curves plotted against the wavenumber for the fixed parameters  $W=10$, $\chi=0.5$, $D_a=0.1$, and $T_a=100$ as the critical total intensity $\mathbb{J}_c$ varies. }
   \label{fig:effofGc.pdf}
 \end{figure*}

In Fig. \ref{fig:effofGc.pdf}, subfigure (a) illustrates the basic state concentration, while subfigure (b) presents the neutral curves plotted against the wavenumber for fixed parameters \( W = 10 \), \( \chi = 0.5 \), \( D_a = 0.1 \), and \( T_a = 100 \) as the critical total intensity \( \mathbb{J}_c \) varies. When \( \mathbb{J}_c = 0.63 \), the maximum concentration is observed at \( z = 0.5 \). However, as \( \mathbb{J}_c \) is increased to \( 0.65 \), the position of maximum concentration shifts towards the upper layer. This shift is accompanied by a reduction in the corresponding critical Rayleigh number, indicating an increase in system instability.

\subsection{Comparison with non-rotating model}
The proposed model is compared with a similar phototactic bioconvection model in a non-rotating medium \cite{rajput2024mathematical} for Taylor number \( T_a=0 \) in this section. This comparison evaluates both models at various values of Darcy number \(D_a\), while keeping other governing parameters, \( V_c = 10, 15, 20 \) and \( \kappa = 0.5, 1.0\), constant. A summary of the critical wavenumber $\lambda^c$ and critical Rayleigh number $R_a^c$ for both models is presented in Table \ref{tab:table7}. Notably, the two models demonstrate excellent agreement across all evaluated values of Darcy number \(D_a\).

\section{Conclusion}
\label{Conclusion}
This study investigates the intricate interactions between phototaxis, rotation, and porous media in triggering bioconvective instability. The model examines how the porous medium, characterized by the Darcy number, impacts the behavior of the suspension under collimated radiation from above, with rotation occurring around vertical axes. We have discussed both stationary and oscillatory solutions in our analysis. Increasing the Darcy number enhances the instability of the system, leading to lower critical Rayleigh numbers and longer bioconvection pattern wavelengths. However, at higher Taylor numbers, the trend reverses, with increased Darcy numbers reducing pattern wavelengths. Rotation stabilizes the bioconvective system by limiting vertical fluid motion and confining dynamics to the horizontal plane. Higher Taylor numbers increase the critical Rayleigh number, suppressing instability. The transition from oscillatory to stationary solutions is observed with increasing Taylor numbers, demonstrating how rotational dynamics influence the nature of instability modes. The analysis of the effect of varying the critical intensity reveals significant changes in the porous medium. An increase in the critical intensity is associated with a reduction in the critical Rayleigh number, which indicates heightened instability within the system. These findings underscore the sensitivity of phototactic bioconvection systems to variations in total intensity, providing deeper insights into their stability characteristics.
 
 This work bridges significant gaps in understanding light-induced bioconvection in rotating porous media, offering theoretical insights applicable to natural and engineered systems. Future research could extend this framework to consider nonlinear stability analyses or the impact of additional environmental factors, such as temperature gradients or chemical reactions. These results not only deepen our understanding of phototactic bioconvection but also pave the way for innovative applications in environmental and industrial processes, such as bioremediation, nutrient transport, and advanced filtration systems.

\nocite{*}
\bibliography{aipsamp}

\providecommand{\noopsort}[1]{}\providecommand{\singleletter}[1]{#1}%
\begin{thebibliography}{43}%
\makeatletter
\providecommand \@ifxundefined [1]{%
 \@ifx{#1\undefined}
}%
\providecommand \@ifnum [1]{%
 \ifnum #1\expandafter \@firstoftwo
 \else \expandafter \@secondoftwo
 \fi
}%
\providecommand \@ifx [1]{%
 \ifx #1\expandafter \@firstoftwo
 \else \expandafter \@secondoftwo
 \fi
}%
\providecommand \natexlab [1]{#1}%
\providecommand \enquote  [1]{``#1''}%
\providecommand \bibnamefont  [1]{#1}%
\providecommand \bibfnamefont [1]{#1}%
\providecommand \citenamefont [1]{#1}%
\providecommand \href@noop [0]{\@secondoftwo}%
\providecommand \href [0]{\begingroup \@sanitize@url \@href}%
\providecommand \@href[1]{\@@startlink{#1}\@@href}%
\providecommand \@@href[1]{\endgroup#1\@@endlink}%
\providecommand \@sanitize@url [0]{\catcode `\\12\catcode `\$12\catcode `\&12\catcode `\#12\catcode `\^12\catcode `\_12\catcode `\%12\relax}%
\providecommand \@@startlink[1]{}%
\providecommand \@@endlink[0]{}%
\providecommand \url  [0]{\begingroup\@sanitize@url \@url }%
\providecommand \@url [1]{\endgroup\@href {#1}{\urlprefix }}%
\providecommand \urlprefix  [0]{URL }%
\providecommand \Eprint [0]{\href }%
\providecommand \doibase [0]{http://dx.doi.org/}%
\providecommand \selectlanguage [0]{\@gobble}%
\providecommand \bibinfo  [0]{\@secondoftwo}%
\providecommand \bibfield  [0]{\@secondoftwo}%
\providecommand \translation [1]{[#1]}%
\providecommand \BibitemOpen [0]{}%
\providecommand \bibitemStop [0]{}%
\providecommand \bibitemNoStop [0]{.\EOS\space}%
\providecommand \EOS [0]{\spacefactor3000\relax}%
\providecommand \BibitemShut  [1]{\csname bibitem#1\endcsname}%
\let\auto@bib@innerbib\@empty
\bibitem [{\citenamefont {Pedley}\ and\ \citenamefont {Kessler}(1992)}]{ref1}%
  \BibitemOpen
  \bibfield  {author} {\bibinfo {author} {\bibfnamefont {T.~J.}\ \bibnamefont {Pedley}}\ and\ \bibinfo {author} {\bibfnamefont {J.~O.}\ \bibnamefont {Kessler}},\ }\bibfield  {title} {\enquote {\bibinfo {title} {Hydrodynamic phenomena in suspensions of swimming micro-organisms},}\ }\href@noop {} {\bibfield  {journal} {\bibinfo  {journal} {Annu. Rev. Fluid Mech.}\ }\textbf {\bibinfo {volume} {24}},\ \bibinfo {pages} {313--358} (\bibinfo {year} {1992})}\BibitemShut {NoStop}%
\bibitem [{\citenamefont {Platt}(1961)}]{ref3}%
  \BibitemOpen
  \bibfield  {author} {\bibinfo {author} {\bibfnamefont {J.~R.}\ \bibnamefont {Platt}},\ }\bibfield  {title} {\enquote {\bibinfo {title} {‘‘bioconvection patterns’’ in cultures of free-swimming organisms},}\ }\href@noop {} {\bibfield  {journal} {\bibinfo  {journal} {Science}\ }\textbf {\bibinfo {volume} {133}},\ \bibinfo {pages} {1766--1767} (\bibinfo {year} {1961})}\BibitemShut {NoStop}%
\bibitem [{\citenamefont {Bees}(2020)}]{ref7a}%
  \BibitemOpen
  \bibfield  {author} {\bibinfo {author} {\bibfnamefont {M.~A.}\ \bibnamefont {Bees}},\ }\bibfield  {title} {\enquote {\bibinfo {title} {Advances in bioconvection},}\ }\href@noop {} {\bibfield  {journal} {\bibinfo  {journal} {Annual Review of Fluid Mechanics}\ }\textbf {\bibinfo {volume} {52}},\ \bibinfo {pages} {449--476} (\bibinfo {year} {2020})}\BibitemShut {NoStop}%
\bibitem [{\citenamefont {Kessler}(985b)}]{ref5}%
  \BibitemOpen
  \bibfield  {author} {\bibinfo {author} {\bibfnamefont {J.~O.}\ \bibnamefont {Kessler}},\ }\bibfield  {title} {\enquote {\bibinfo {title} {Co-operative and concentrative phenomena of swimming microorganisms},}\ }\href@noop {} {\bibfield  {journal} {\bibinfo  {journal} {Contemp. Phys.}\ }\textbf {\bibinfo {volume} {26}},\ \bibinfo {pages} {147--166} (\bibinfo {year} {1985b})}\BibitemShut {NoStop}%
\bibitem [{\citenamefont {Wager}(1911)}]{ref2}%
  \BibitemOpen
  \bibfield  {author} {\bibinfo {author} {\bibfnamefont {H.~W.~T.}\ \bibnamefont {Wager}},\ }\bibfield  {title} {\enquote {\bibinfo {title} {On the effect of gravity upon the movements and aggregation of euglena viridis, ehrb., and other micro-organisms},}\ }\href@noop {} {\bibfield  {journal} {\bibinfo  {journal} {Phil. Trans. R. Soc. Lond. B}\ }\textbf {\bibinfo {volume} {201}},\ \bibinfo {pages} {333--390} (\bibinfo {year} {1911})}\BibitemShut {NoStop}%
\bibitem [{\citenamefont {Nultsch}\ and\ \citenamefont {Hoff}(1993)}]{ref4}%
  \BibitemOpen
  \bibfield  {author} {\bibinfo {author} {\bibfnamefont {W.}~\bibnamefont {Nultsch}}\ and\ \bibinfo {author} {\bibfnamefont {E.}~\bibnamefont {Hoff}},\ }\bibfield  {title} {\enquote {\bibinfo {title} {Investigations on pattern formatin in euglenae},}\ }\href@noop {} {\bibfield  {journal} {\bibinfo  {journal} {Arch. Protistenk}\ }\textbf {\bibinfo {volume} {115}},\ \bibinfo {pages} {336--352} (\bibinfo {year} {1993})}\BibitemShut {NoStop}%
\bibitem [{\citenamefont {Brinkmann}(1968)}]{ref6}%
  \BibitemOpen
  \bibfield  {author} {\bibinfo {author} {\bibfnamefont {K.}~\bibnamefont {Brinkmann}},\ }\bibfield  {title} {\enquote {\bibinfo {title} {An phasengrenzen induzierte ein und zweidimensionale kristallmuster in kulturen von euglena gracilis},}\ }\href@noop {} {\bibfield  {journal} {\bibinfo  {journal} {Z. Pflanzen Physiol.}\ }\textbf {\bibinfo {volume} {59}},\ \bibinfo {pages} {364--376} (\bibinfo {year} {1968})}\BibitemShut {NoStop}%
\bibitem [{\citenamefont {Williams}\ and\ \citenamefont {Bees}(2011)}]{ref7}%
  \BibitemOpen
  \bibfield  {author} {\bibinfo {author} {\bibfnamefont {C.~R.}\ \bibnamefont {Williams}}\ and\ \bibinfo {author} {\bibfnamefont {M.~A.}\ \bibnamefont {Bees}},\ }\bibfield  {title} {\enquote {\bibinfo {title} {A tale of three taxes: Photo-gyro-gravitactic bioconvection},}\ }\href@noop {} {\bibfield  {journal} {\bibinfo  {journal} {J. Exp. Biol.}\ }\textbf {\bibinfo {volume} {214}},\ \bibinfo {pages} {2398--2408} (\bibinfo {year} {2011})}\BibitemShut {NoStop}%
\bibitem [{\citenamefont {Häder}(1987)}]{ref8}%
  \BibitemOpen
  \bibfield  {author} {\bibinfo {author} {\bibfnamefont {D.~P.}\ \bibnamefont {Häder}},\ }\bibfield  {title} {\enquote {\bibinfo {title} {Polarotaxis, gravitaxis and vertical phototaxis in the green flagellate, euglena gracilis},}\ }\href@noop {} {\bibfield  {journal} {\bibinfo  {journal} {Arch. Microbiol.}\ }\textbf {\bibinfo {volume} {147}},\ \bibinfo {pages} {179--183} (\bibinfo {year} {1987})}\BibitemShut {NoStop}%
\bibitem [{\citenamefont {Kopp}\ and\ \citenamefont {Yanovsky}(2023{\natexlab{a}})}]{kopp2023effect}%
  \BibitemOpen
  \bibfield  {author} {\bibinfo {author} {\bibfnamefont {M.}~\bibnamefont {Kopp}}\ and\ \bibinfo {author} {\bibfnamefont {V.}~\bibnamefont {Yanovsky}},\ }\bibfield  {title} {\enquote {\bibinfo {title} {Effect of gravity modulation on weakly nonlinear bio-thermal convection in a porous medium layer},}\ }\href@noop {} {\bibfield  {journal} {\bibinfo  {journal} {Journal of Applied Physics}\ }\textbf {\bibinfo {volume} {134}} (\bibinfo {year} {2023}{\natexlab{a}})}\BibitemShut {NoStop}%
\bibitem [{\citenamefont {Rajput}\ and\ \citenamefont {Panda}(2024{\natexlab{a}})}]{rajput2024mathematical}%
  \BibitemOpen
  \bibfield  {author} {\bibinfo {author} {\bibfnamefont {S.}~\bibnamefont {Rajput}}\ and\ \bibinfo {author} {\bibfnamefont {M.}~\bibnamefont {Panda}},\ }\bibfield  {title} {\enquote {\bibinfo {title} {A mathematical modeling of light-induced bioconvection in an isotropic porous medium},}\ }\href@noop {} {\bibfield  {journal} {\bibinfo  {journal} {Chinese Journal of Physics}\ }\textbf {\bibinfo {volume} {91}},\ \bibinfo {pages} {792--806} (\bibinfo {year} {2024}{\natexlab{a}})}\BibitemShut {NoStop}%
\bibitem [{\citenamefont {Kuznetsov}\ and\ \citenamefont {Jiang}(2001)}]{kuznetsov2001numerical}%
  \BibitemOpen
  \bibfield  {author} {\bibinfo {author} {\bibfnamefont {A.}~\bibnamefont {Kuznetsov}}\ and\ \bibinfo {author} {\bibfnamefont {N.}~\bibnamefont {Jiang}},\ }\bibfield  {title} {\enquote {\bibinfo {title} {Numerical investigation of bioconvection of gravitactic microorganisms in an isotropic porous medium},}\ }\href@noop {} {\bibfield  {journal} {\bibinfo  {journal} {International communications in heat and mass transfer}\ }\textbf {\bibinfo {volume} {28}},\ \bibinfo {pages} {877--886} (\bibinfo {year} {2001})}\BibitemShut {NoStop}%
\bibitem [{\citenamefont {Kuznetsov}\ and\ \citenamefont {Avramenko}(2003)}]{kuznetsov2003stability}%
  \BibitemOpen
  \bibfield  {author} {\bibinfo {author} {\bibfnamefont {A.}~\bibnamefont {Kuznetsov}}\ and\ \bibinfo {author} {\bibfnamefont {A.}~\bibnamefont {Avramenko}},\ }\bibfield  {title} {\enquote {\bibinfo {title} {Stability analysis of bioconvection of gyrotactic motile microorganisms in a fluid saturated porous medium},}\ }\href@noop {} {\bibfield  {journal} {\bibinfo  {journal} {Transport in porous media}\ }\textbf {\bibinfo {volume} {53}},\ \bibinfo {pages} {95--104} (\bibinfo {year} {2003})}\BibitemShut {NoStop}%
\bibitem [{\citenamefont {Nield}, \citenamefont {Kuznetsov},\ and\ \citenamefont {Avramenko}(2004)}]{nield2004onset}%
  \BibitemOpen
  \bibfield  {author} {\bibinfo {author} {\bibfnamefont {D.}~\bibnamefont {Nield}}, \bibinfo {author} {\bibfnamefont {A.}~\bibnamefont {Kuznetsov}}, \ and\ \bibinfo {author} {\bibfnamefont {A.}~\bibnamefont {Avramenko}},\ }\bibfield  {title} {\enquote {\bibinfo {title} {The onset of bioconvection in a horizontal porous-medium layer},}\ }\href@noop {} {\bibfield  {journal} {\bibinfo  {journal} {Transport in porous media}\ }\textbf {\bibinfo {volume} {54}},\ \bibinfo {pages} {335--344} (\bibinfo {year} {2004})}\BibitemShut {NoStop}%
\bibitem [{\citenamefont {Avramenko}\ and\ \citenamefont {Kuznetsov}(2006)}]{avramenko2006onset}%
  \BibitemOpen
  \bibfield  {author} {\bibinfo {author} {\bibfnamefont {A.}~\bibnamefont {Avramenko}}\ and\ \bibinfo {author} {\bibfnamefont {A.}~\bibnamefont {Kuznetsov}},\ }\bibfield  {title} {\enquote {\bibinfo {title} {The onset of convection in a suspension of gyrotactic microorganisms in superimposed fluid and porous layers: effect of vertical throughflow},}\ }\href@noop {} {\bibfield  {journal} {\bibinfo  {journal} {Transport in porous media}\ }\textbf {\bibinfo {volume} {65}},\ \bibinfo {pages} {159--176} (\bibinfo {year} {2006})}\BibitemShut {NoStop}%
\bibitem [{\citenamefont {Biswas}\ \emph {et~al.}(2020)\citenamefont {Biswas}, \citenamefont {Datta}, \citenamefont {Manna}, \citenamefont {Mandal},\ and\ \citenamefont {Gorla}}]{biswas2020thermo}%
  \BibitemOpen
  \bibfield  {author} {\bibinfo {author} {\bibfnamefont {N.}~\bibnamefont {Biswas}}, \bibinfo {author} {\bibfnamefont {A.}~\bibnamefont {Datta}}, \bibinfo {author} {\bibfnamefont {N.~K.}\ \bibnamefont {Manna}}, \bibinfo {author} {\bibfnamefont {D.~K.}\ \bibnamefont {Mandal}}, \ and\ \bibinfo {author} {\bibfnamefont {R.~S.~R.}\ \bibnamefont {Gorla}},\ }\bibfield  {title} {\enquote {\bibinfo {title} {Thermo-bioconvection of oxytactic microorganisms in porous media in the presence of magnetic field},}\ }\href@noop {} {\bibfield  {journal} {\bibinfo  {journal} {International Journal of Numerical Methods for Heat \& Fluid Flow}\ }\textbf {\bibinfo {volume} {31}},\ \bibinfo {pages} {1638--1661} (\bibinfo {year} {2020})}\BibitemShut {NoStop}%
\bibitem [{\citenamefont {Kopp}\ and\ \citenamefont {Yanovsky}(2023{\natexlab{b}})}]{kopp2023darcy}%
  \BibitemOpen
  \bibfield  {author} {\bibinfo {author} {\bibfnamefont {M.}~\bibnamefont {Kopp}}\ and\ \bibinfo {author} {\bibfnamefont {V.}~\bibnamefont {Yanovsky}},\ }\bibfield  {title} {\enquote {\bibinfo {title} {Darcy--brinkman bio-thermal convection in a porous rotating layer saturated by a newtonian fluid containing gyrotactic microorganisms},}\ }\href@noop {} {\bibfield  {journal} {\bibinfo  {journal} {Ukrainian journal of physics}\ }\textbf {\bibinfo {volume} {68}},\ \bibinfo {pages} {30--30} (\bibinfo {year} {2023}{\natexlab{b}})}\BibitemShut {NoStop}%
\bibitem [{\citenamefont {Kopp}\ and\ \citenamefont {Yanovsky}(2024)}]{kopp2024weakly}%
  \BibitemOpen
  \bibfield  {author} {\bibinfo {author} {\bibfnamefont {M.~I.}\ \bibnamefont {Kopp}}\ and\ \bibinfo {author} {\bibfnamefont {V.~V.}\ \bibnamefont {Yanovsky}},\ }\bibfield  {title} {\enquote {\bibinfo {title} {Weakly nonlinear bio-thermal convection in a porous media layer under rotation, gravity modulation, and heat source},}\ }\href@noop {} {\bibfield  {journal} {\bibinfo  {journal} {East European Journal of Physics}\ ,\ \bibinfo {pages} {175--191}} (\bibinfo {year} {2024})}\BibitemShut {NoStop}%
\bibitem [{\citenamefont {Greenspan}(1968)}]{ref11}%
  \BibitemOpen
  \bibfield  {author} {\bibinfo {author} {\bibfnamefont {H.~P.}\ \bibnamefont {Greenspan}},\ }\enquote {\bibinfo {title} {\textit{The theory of rotating fluids}},}\ \ (\bibinfo  {publisher} {Cambridge University Press},\ \bibinfo {address} {London},\ \bibinfo {year} {1968})\BibitemShut {NoStop}%
\bibitem [{\citenamefont {Chandrasekhar}(1961)}]{ref12}%
  \BibitemOpen
  \bibfield  {author} {\bibinfo {author} {\bibfnamefont {S.}~\bibnamefont {Chandrasekhar}},\ }\enquote {\bibinfo {title} {\textit{Hydrodynamic and hydromagnetic stability}},}\ \ (\bibinfo  {publisher} {Oxford University Press},\ \bibinfo {year} {1961})\BibitemShut {NoStop}%
\bibitem [{\citenamefont {Sekar}\ and\ \citenamefont {Vaidyanathan}(1993)}]{ref12-1}%
  \BibitemOpen
  \bibfield  {author} {\bibinfo {author} {\bibfnamefont {R.}~\bibnamefont {Sekar}}\ and\ \bibinfo {author} {\bibfnamefont {G.}~\bibnamefont {Vaidyanathan}},\ }\bibfield  {title} {\enquote {\bibinfo {title} {Convective instability of a magnetized ferrofluid in a rotating porous medium},}\ }\href@noop {} {\bibfield  {journal} {\bibinfo  {journal} {International journal of engineering science}\ }\textbf {\bibinfo {volume} {31}},\ \bibinfo {pages} {1139--1150} (\bibinfo {year} {1993})}\BibitemShut {NoStop}%
\bibitem [{\citenamefont {Venkatasubramanian}\ and\ \citenamefont {Kaloni}(1994)}]{ref12-2}%
  \BibitemOpen
  \bibfield  {author} {\bibinfo {author} {\bibfnamefont {S.}~\bibnamefont {Venkatasubramanian}}\ and\ \bibinfo {author} {\bibfnamefont {P.}~\bibnamefont {Kaloni}},\ }\bibfield  {title} {\enquote {\bibinfo {title} {Effects of rotation on the thermoconvective instability of a horizontal layer of ferrofluids},}\ }\href@noop {} {\bibfield  {journal} {\bibinfo  {journal} {International journal of engineering science}\ }\textbf {\bibinfo {volume} {32}},\ \bibinfo {pages} {237--256} (\bibinfo {year} {1994})}\BibitemShut {NoStop}%
\bibitem [{\citenamefont {Auernhammer}\ and\ \citenamefont {Brand}(2000)}]{ref12-3}%
  \BibitemOpen
  \bibfield  {author} {\bibinfo {author} {\bibfnamefont {G.}~\bibnamefont {Auernhammer}}\ and\ \bibinfo {author} {\bibfnamefont {H.}~\bibnamefont {Brand}},\ }\bibfield  {title} {\enquote {\bibinfo {title} {Thermal convection in a rotating layer of a magnetic fluid},}\ }\href@noop {} {\bibfield  {journal} {\bibinfo  {journal} {The European Physical Journal B-Condensed Matter and Complex Systems}\ }\textbf {\bibinfo {volume} {16}},\ \bibinfo {pages} {157--168} (\bibinfo {year} {2000})}\BibitemShut {NoStop}%
\bibitem [{\citenamefont {Ruo}, \citenamefont {Chang},\ and\ \citenamefont {Chen}(2010)}]{ref12-4}%
  \BibitemOpen
  \bibfield  {author} {\bibinfo {author} {\bibfnamefont {A.-C.}\ \bibnamefont {Ruo}}, \bibinfo {author} {\bibfnamefont {M.-H.}\ \bibnamefont {Chang}}, \ and\ \bibinfo {author} {\bibfnamefont {F.}~\bibnamefont {Chen}},\ }\bibfield  {title} {\enquote {\bibinfo {title} {Effect of rotation on the electrohydrodynamic instability of a fluid layer with an electrical conductivity gradient},}\ }\href@noop {} {\bibfield  {journal} {\bibinfo  {journal} {Physics of Fluids}\ }\textbf {\bibinfo {volume} {22}},\ \bibinfo {pages} {024102} (\bibinfo {year} {2010})}\BibitemShut {NoStop}%
\bibitem [{\citenamefont {Yadav}, \citenamefont {Agrawal},\ and\ \citenamefont {Bhargava}(2011)}]{ref12-5}%
  \BibitemOpen
  \bibfield  {author} {\bibinfo {author} {\bibfnamefont {D.}~\bibnamefont {Yadav}}, \bibinfo {author} {\bibfnamefont {G.}~\bibnamefont {Agrawal}}, \ and\ \bibinfo {author} {\bibfnamefont {R.}~\bibnamefont {Bhargava}},\ }\bibfield  {title} {\enquote {\bibinfo {title} {Thermal instability of rotating nanofluid layer},}\ }\href@noop {} {\bibfield  {journal} {\bibinfo  {journal} {International Journal of Engineering Science}\ }\textbf {\bibinfo {volume} {49}},\ \bibinfo {pages} {1171--1184} (\bibinfo {year} {2011})}\BibitemShut {NoStop}%
\bibitem [{\citenamefont {Mahajan}\ and\ \citenamefont {Arora}(2013)}]{ref12-6}%
  \BibitemOpen
  \bibfield  {author} {\bibinfo {author} {\bibfnamefont {A.}~\bibnamefont {Mahajan}}\ and\ \bibinfo {author} {\bibfnamefont {M.}~\bibnamefont {Arora}},\ }\bibfield  {title} {\enquote {\bibinfo {title} {Convection in rotating magnetic nanofluids},}\ }\href@noop {} {\bibfield  {journal} {\bibinfo  {journal} {Applied Mathematics and Computation}\ }\textbf {\bibinfo {volume} {219}},\ \bibinfo {pages} {6284--6296} (\bibinfo {year} {2013})}\BibitemShut {NoStop}%
\bibitem [{\citenamefont {Waqas}\ \emph {et~al.}(2021)\citenamefont {Waqas}, \citenamefont {Naseem}, \citenamefont {Muhammad},\ and\ \citenamefont {Farooq}}]{waqas}%
  \BibitemOpen
  \bibfield  {author} {\bibinfo {author} {\bibfnamefont {H.}~\bibnamefont {Waqas}}, \bibinfo {author} {\bibfnamefont {R.}~\bibnamefont {Naseem}}, \bibinfo {author} {\bibfnamefont {T.}~\bibnamefont {Muhammad}}, \ and\ \bibinfo {author} {\bibfnamefont {U.}~\bibnamefont {Farooq}},\ }\bibfield  {title} {\enquote {\bibinfo {title} {Bioconvection flow of casson nanofluid by rotating disk with motile microorganisms},}\ }\href@noop {} {\bibfield  {journal} {\bibinfo  {journal} {Journal of Materials Research and Technology}\ }\textbf {\bibinfo {volume} {13}},\ \bibinfo {pages} {2392--2407} (\bibinfo {year} {2021})}\BibitemShut {NoStop}%
\bibitem [{\citenamefont {Vincent}\ and\ \citenamefont {Hill}(1996)}]{ref9}%
  \BibitemOpen
  \bibfield  {author} {\bibinfo {author} {\bibfnamefont {R.~V.}\ \bibnamefont {Vincent}}\ and\ \bibinfo {author} {\bibfnamefont {N.~A.}\ \bibnamefont {Hill}},\ }\bibfield  {title} {\enquote {\bibinfo {title} {Bioconvection in a suspension of phototactic algae},}\ }\href@noop {} {\bibfield  {journal} {\bibinfo  {journal} {J. Fluid Mech.}\ }\textbf {\bibinfo {volume} {327}},\ \bibinfo {pages} {343--371} (\bibinfo {year} {1996})}\BibitemShut {NoStop}%
\bibitem [{\citenamefont {Ghorai}\ and\ \citenamefont {Hill}(2005)}]{ref13}%
  \BibitemOpen
  \bibfield  {author} {\bibinfo {author} {\bibfnamefont {S.}~\bibnamefont {Ghorai}}\ and\ \bibinfo {author} {\bibfnamefont {N.~A.}\ \bibnamefont {Hill}},\ }\bibfield  {title} {\enquote {\bibinfo {title} {Penetrative phototactic bioconvection},}\ }\href@noop {} {\bibfield  {journal} {\bibinfo  {journal} {Phys. Fluids}\ }\textbf {\bibinfo {volume} {17}},\ \bibinfo {pages} {074101} (\bibinfo {year} {2005})}\BibitemShut {NoStop}%
\bibitem [{\citenamefont {Ghorai}, \citenamefont {Panda},\ and\ \citenamefont {Hill}(2010)}]{ref14}%
  \BibitemOpen
  \bibfield  {author} {\bibinfo {author} {\bibfnamefont {S.}~\bibnamefont {Ghorai}}, \bibinfo {author} {\bibfnamefont {M.~K.}\ \bibnamefont {Panda}}, \ and\ \bibinfo {author} {\bibfnamefont {N.~A.}\ \bibnamefont {Hill}},\ }\bibfield  {title} {\enquote {\bibinfo {title} {Bioconvection in a suspension of isotropically scattering phototactic algae},}\ }\href@noop {} {\bibfield  {journal} {\bibinfo  {journal} {Phys. Fluids}\ }\textbf {\bibinfo {volume} {22}},\ \bibinfo {pages} {071901} (\bibinfo {year} {2010})}\BibitemShut {NoStop}%
\bibitem [{\citenamefont {Kumar}(2023{\natexlab{a}})}]{kumar2023}%
  \BibitemOpen
  \bibfield  {author} {\bibinfo {author} {\bibfnamefont {S.}~\bibnamefont {Kumar}},\ }\bibfield  {title} {\enquote {\bibinfo {title} {Isotropic scattering with a rigid upper surface at the onset of phototactic bioconvection},}\ }\href@noop {} {\bibfield  {journal} {\bibinfo  {journal} {Physics of Fluids}\ }\textbf {\bibinfo {volume} {35}},\ \bibinfo {pages} {024106} (\bibinfo {year} {2023}{\natexlab{a}})}\BibitemShut {NoStop}%
\bibitem [{\citenamefont {Ghorai}\ and\ \citenamefont {Panda}(2013)}]{ref15}%
  \BibitemOpen
  \bibfield  {author} {\bibinfo {author} {\bibfnamefont {S.}~\bibnamefont {Ghorai}}\ and\ \bibinfo {author} {\bibfnamefont {M.~K.}\ \bibnamefont {Panda}},\ }\bibfield  {title} {\enquote {\bibinfo {title} {Bioconvection in an anisotropic scattering suspension of phototactic algae},}\ }\href@noop {} {\bibfield  {journal} {\bibinfo  {journal} {Eur. J. Mech.-B/Fluids}\ }\textbf {\bibinfo {volume} {41}},\ \bibinfo {pages} {81--93} (\bibinfo {year} {2013})}\BibitemShut {NoStop}%
\bibitem [{\citenamefont {Panda}\ \emph {et~al.}(2016)\citenamefont {Panda}, \citenamefont {Singh}, \citenamefont {Mishra},\ and\ \citenamefont {Mohanty}}]{ref16}%
  \BibitemOpen
  \bibfield  {author} {\bibinfo {author} {\bibfnamefont {M.~K.}\ \bibnamefont {Panda}}, \bibinfo {author} {\bibfnamefont {R.}~\bibnamefont {Singh}}, \bibinfo {author} {\bibfnamefont {A.~C.}\ \bibnamefont {Mishra}}, \ and\ \bibinfo {author} {\bibfnamefont {S.~K.}\ \bibnamefont {Mohanty}},\ }\bibfield  {title} {\enquote {\bibinfo {title} {Effects of both diffuse and collimated incident radiation on phototactic bioconvection},}\ }\href@noop {} {\bibfield  {journal} {\bibinfo  {journal} {Phys. Fluids}\ }\textbf {\bibinfo {volume} {28}},\ \bibinfo {pages} {124104} (\bibinfo {year} {2016})}\BibitemShut {NoStop}%
\bibitem [{\citenamefont {Panda}, \citenamefont {Sharma},\ and\ \citenamefont {Kumar}(2022)}]{ref17}%
  \BibitemOpen
  \bibfield  {author} {\bibinfo {author} {\bibfnamefont {M.~K.}\ \bibnamefont {Panda}}, \bibinfo {author} {\bibfnamefont {P.}~\bibnamefont {Sharma}}, \ and\ \bibinfo {author} {\bibfnamefont {S.}~\bibnamefont {Kumar}},\ }\bibfield  {title} {\enquote {\bibinfo {title} {Effect of oblique irradiation on the onset of phototactic bioconvection},}\ }\href@noop {} {\bibfield  {journal} {\bibinfo  {journal} {Phys. Fluids}\ }\textbf {\bibinfo {volume} {34}},\ \bibinfo {pages} {024108} (\bibinfo {year} {2022})}\BibitemShut {NoStop}%
\bibitem [{\citenamefont {Kumar}(2022)}]{ref18}%
  \BibitemOpen
  \bibfield  {author} {\bibinfo {author} {\bibfnamefont {S.}~\bibnamefont {Kumar}},\ }\bibfield  {title} {\enquote {\bibinfo {title} {Phototactic isotropic scattering bioconvection with oblique irradiation},}\ }\href@noop {} {\bibfield  {journal} {\bibinfo  {journal} {Phys. Fluids}\ }\textbf {\bibinfo {volume} {34}},\ \bibinfo {pages} {114125} (\bibinfo {year} {2022})}\BibitemShut {NoStop}%
\bibitem [{\citenamefont {Kumar}(2023{\natexlab{b}})}]{kumar2023effect}%
  \BibitemOpen
  \bibfield  {author} {\bibinfo {author} {\bibfnamefont {S.}~\bibnamefont {Kumar}},\ }\bibfield  {title} {\enquote {\bibinfo {title} {Effect of rotation on the suspension of phototactic bioconvection},}\ }\href@noop {} {\bibfield  {journal} {\bibinfo  {journal} {Physics of Fluids}\ }\textbf {\bibinfo {volume} {35}} (\bibinfo {year} {2023}{\natexlab{b}})}\BibitemShut {NoStop}%
\bibitem [{\citenamefont {Rajput}\ and\ \citenamefont {Panda}(2024{\natexlab{b}})}]{rajput2024effect}%
  \BibitemOpen
  \bibfield  {author} {\bibinfo {author} {\bibfnamefont {S.}~\bibnamefont {Rajput}}\ and\ \bibinfo {author} {\bibfnamefont {M.}~\bibnamefont {Panda}},\ }\bibfield  {title} {\enquote {\bibinfo {title} {Effect of scattered/diffuse flux on the phototactic bioconvection in the absence of collimated flux},}\ }\href@noop {} {\bibfield  {journal} {\bibinfo  {journal} {Physics of Fluids}\ }\textbf {\bibinfo {volume} {36}} (\bibinfo {year} {2024}{\natexlab{b}})}\BibitemShut {NoStop}%
\bibitem [{\citenamefont {Kumar}\ and\ \citenamefont {Wang}(2024{\natexlab{a}})}]{kumar2024thermal}%
  \BibitemOpen
  \bibfield  {author} {\bibinfo {author} {\bibfnamefont {S.}~\bibnamefont {Kumar}}\ and\ \bibinfo {author} {\bibfnamefont {S.}~\bibnamefont {Wang}},\ }\bibfield  {title} {\enquote {\bibinfo {title} {Thermal-bioconvection instability in suspensions of phototactic microorganisms heated from below},}\ }\href@noop {} {\bibfield  {journal} {\bibinfo  {journal} {Thermal Science and Engineering Progress}\ ,\ \bibinfo {pages} {102691}} (\bibinfo {year} {2024}{\natexlab{a}})}\BibitemShut {NoStop}%
\bibitem [{\citenamefont {Kumar}\ and\ \citenamefont {Wang}(2024{\natexlab{b}})}]{sandeep2024thermal}%
  \BibitemOpen
  \bibfield  {author} {\bibinfo {author} {\bibfnamefont {S.}~\bibnamefont {Kumar}}\ and\ \bibinfo {author} {\bibfnamefont {S.}~\bibnamefont {Wang}},\ }\bibfield  {title} {\enquote {\bibinfo {title} {Heating from above in non-scattering suspension: Phototactic bioconvection under collimated irradiation},}\ }\href@noop {} {\bibfield  {journal} {\bibinfo  {journal} {Physics of Fluids}\ }\textbf {\bibinfo {volume} {36}} (\bibinfo {year} {2024}{\natexlab{b}})}\BibitemShut {NoStop}%
\bibitem [{\citenamefont {Modest}(2003)}]{ref-modest}%
  \BibitemOpen
  \bibfield  {author} {\bibinfo {author} {\bibfnamefont {M.~F.}\ \bibnamefont {Modest}},\ }\enquote {\bibinfo {title} {\textit{Radiative Heat Transfer}},}\ \ (\bibinfo  {publisher} {Academic Press},\ \bibinfo {address} {New York},\ \bibinfo {year} {2003})\ \bibinfo {edition} {2nd}\ ed.\BibitemShut {Stop}%
\bibitem [{\citenamefont {Chandrasekhar}(1960)}]{ref-chand}%
  \BibitemOpen
  \bibfield  {author} {\bibinfo {author} {\bibfnamefont {S.}~\bibnamefont {Chandrasekhar}},\ }\enquote {\bibinfo {title} {\textit{Radiative Transfer}},}\ \ (\bibinfo  {publisher} {Dover},\ \bibinfo {address} {New York},\ \bibinfo {year} {1960})\BibitemShut {NoStop}%
\bibitem [{\citenamefont {Shampine}, \citenamefont {Gladwell},\ and\ \citenamefont {Thompson}(2003)}]{shampine2003solving}%
  \BibitemOpen
  \bibfield  {author} {\bibinfo {author} {\bibfnamefont {L.~F.}\ \bibnamefont {Shampine}}, \bibinfo {author} {\bibfnamefont {I.}~\bibnamefont {Gladwell}}, \ and\ \bibinfo {author} {\bibfnamefont {S.}~\bibnamefont {Thompson}},\ }\href@noop {} {\emph {\bibinfo {title} {Solving ODEs with matlab}}}\ (\bibinfo  {publisher} {Cambridge university press},\ \bibinfo {year} {2003})\BibitemShut {NoStop}%
\bibitem [{\citenamefont {Chandrasekhar}(1953)}]{chand1953}%
  \BibitemOpen
  \bibfield  {author} {\bibinfo {author} {\bibfnamefont {S.}~\bibnamefont {Chandrasekhar}},\ }\bibfield  {title} {\enquote {\bibinfo {title} {The instability of a layer of fluid heated below and subject to coriolis forces},}\ }\href@noop {} {\bibfield  {journal} {\bibinfo  {journal} {Proceedings of the Royal Society of London. Series A. Mathematical and Physical Sciences}\ }\textbf {\bibinfo {volume} {217}},\ \bibinfo {pages} {306--327} (\bibinfo {year} {1953})}\BibitemShut {NoStop}%
\end{thebibliography}%

\end{document}